\documentstyle[aps,epsfig]{revtex}
\newcommand \beq{\begin{eqnarray}}
\newcommand \eeq{\end{eqnarray}}
\newcommand \ga{\raisebox{-.5ex}{$\stackrel{>}{\sim}$}}
\newcommand \la{\raisebox{-.5ex}{$\stackrel{<}{\sim}$}}
\newcommand \eps{\varepsilon}
\psfull
\begin{document}
\bibliographystyle{unsrt}
\baselineskip=15pt
\title{\bf
    The physics of Hanbury Brown--Twiss intensity interferometry:  from
stars to nuclear collisions.\footnote{Lectures given at the XXXVII Zakopane
School, June 1997.  To be published in Acta Physica Polonica.}
}
\author{
   Gordon Baym\\
   Department of Physics\\
    University of Illinois at Urbana-Champaign,\\
          1110 W. Green St., Urbana, IL 61801, U.S.A.\\}
\maketitle

\begin{abstract}

    In the 1950's Hanbury Brown and Twiss showed that one could measure the
angular sizes of astronomical radio sources and stars from correlations of
signal intensities, rather than amplitudes, in independent detectors.  Their
subsequent correlation experiments demonstrating quantum bunching of photons
in incoherent light beams were seminal in the development of quantum optics.
Since that time the technique of ``intensity interferometry" has become a
valuable probe of high energy nuclear and particle collisions, providing
information on the space-time geometry of the collision.  The effect is one of
the few measurements in elementary particle detection that depends on the wave
mechanics of the produced particles.  Here we discuss the basic physics of
intensity interferometry, and its current applications in high energy nuclear
physics, as well as recent applications in condensed matter and atomic
physics.

\end{abstract}

\section{Introduction}

    Hanbury Brown--Twiss (HBT) interferometry, the measurement of two
identical particle correlations, has become a very important technique in
particle and heavy-ion collisions, enabling one to probe the evolving geometry
of the collision volume.  Figures 1 and 2 illustrate the general idea of an
HBT measurement:  plotted in Fig. 1 is the two-particle correlation function,
$C(Q_{\rm inv})$ -- measured for pairs of $\pi^+$ as well as for pairs of
$\pi^-$ by NA44 for 200 GeV/A S on Pb at the CERN SPS\cite{NA44SPb} -- as a
function of the invariant momentum difference $Q_{\rm inv}=
[(p_1-p_2)^2]^{1/2}$ of the two particles.  The characteristic falloff
distance $\Delta q$ in momentum of the correlation function is of order 50
MeV/c for pions; the length $\hbar/\Delta q$, which is $\sim$ 4 fm, is
basically a measure of the size of the source of the final state pions, the
size of the source when the pions no longer interact strongly with other
particles.  Also shown in Fig. 1, for comparison, is the correlation function
for pairs of $\pi^+$ for 450 GeV protons on Pb, which, being broader,
indicates a smaller source size.  Figure 2 similarly shows the correlation
function for $\pi^+\pi^+$, $\pi^-\pi^-$, and K$^+$K$^+$ pairs produced in
collisions of Au on Au at 10.8 GeV/A measured by E877 at the AGS in
Brookhaven, also as a function of the invariant momentum difference
\cite{e877c}.

    In general, the two-correlation function is defined by
\beq
C(q)= \frac{\{\langle n_{\vec p_1}n_{\vec p_1}\rangle\}}{\{\langle
n_{\vec p_1}\rangle\langle n_{\vec p_2}\rangle\}},
\label{cq}
\eeq
where $n_{\vec p}$ is the number of particles of momentum ${\vec p}$
measured in a single event, $\vec q ={\vec p_1}-{\vec p_2\,}$, and the
averages, denoted by $\langle\cdots\rangle$, are over an ensemble of events.
The pairs in the numerator are taken from the same event, and the pairs in the
denominator from different events.  Usually, one also averages the numerator
and denominator separately over a range of center-of-mass momenta $\vec P =
{\vec p_1}+{\vec p_2}$ of the pair, an average denoted here by $\{\cdots\}$.
As $q$ becomes very large the correlations between the particles are lost, and
the correlation function approaches unity.

    The basic issue I want to discuss in these lectures is how and why HBT
interferometry works.  The effect is in a unique class of experiments
involving multiparticle correlations that are sensitive to the actual wave
mechanics of particles as they stream out to the detectors.  Normally, one
imagines quantum mechanics as being important in high energy experiments only
until the particles leave the interaction region; from then on one usually
pictures them as little bullets on classical trajectories.  (Quantum phenomena
such as kaon regeneration and neutrino oscillations involve the internal
degrees of freedom of the particles, rather than spatial amplitudes.)
Considering the wave mechanics of the emitted particles in space and time is
crucial to understanding questions such as why {\it independent} particle
detectors give a greater signal when they are close together, corresponding to
small $q$, than far apart.  Further issues are:  How accurately are the
distances that are determined by the falloff of the correlation function
related to the size of the system?  In principal the correlation function at
small momenta differences should rise up to 2 for a perfectly chaotic source.
However, it only goes up to $\sim$1.5 - 1.6 for the pion pairs shown in Figs.
1 and 2. What is the physics that reduces the correlation function at small
momentum differences?  What is the effect of final state Coulomb interactions
on the measured correlations?

\begin{figure}
\begin{center}
\epsfig{file=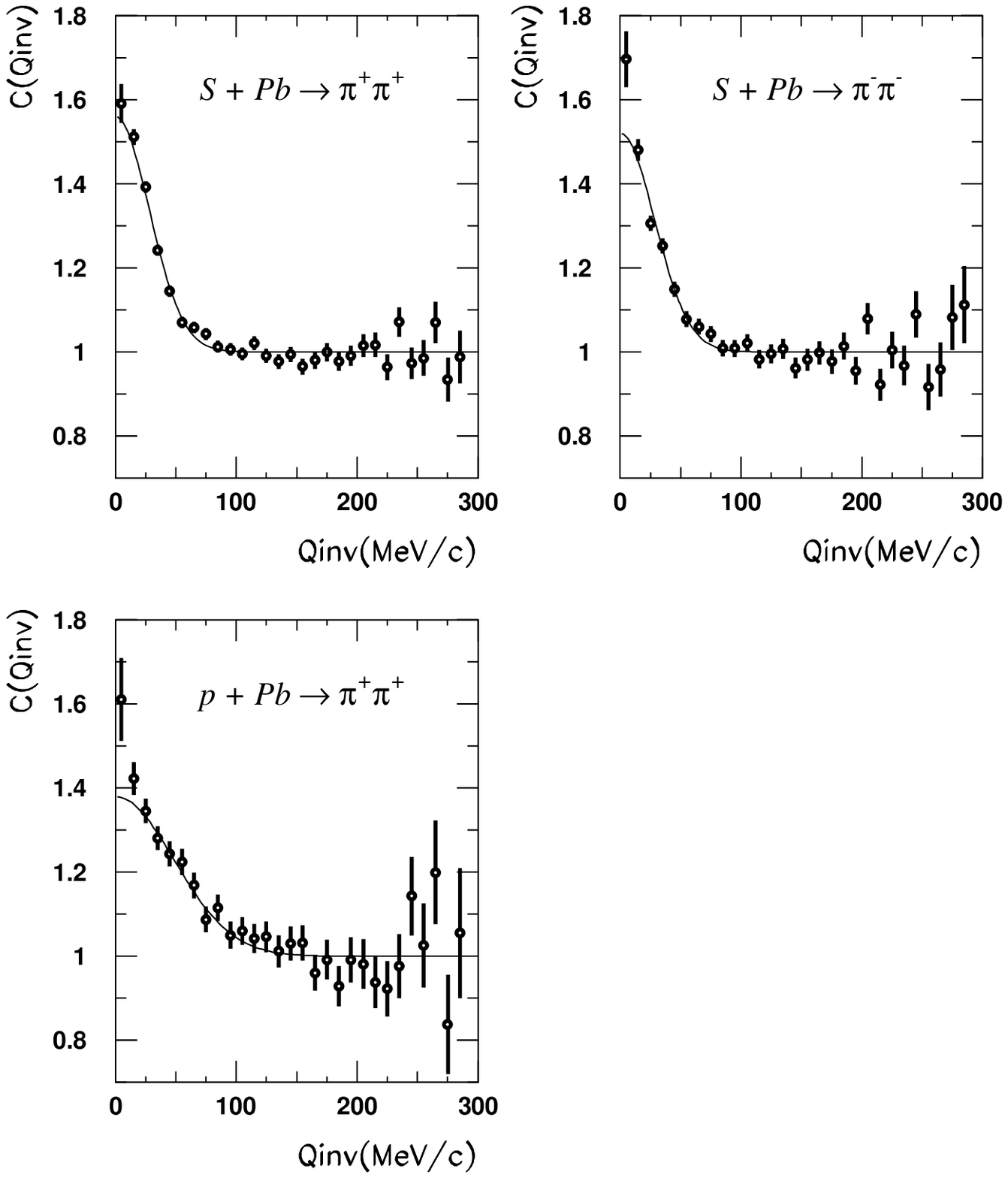, height=12cm}
\end{center}
Fig.~1.  Two-particle correlation function for $\pi^+ \pi^+$ and $\pi^- \pi^-$
pairs in 200 GeV/A collisions of S on Pb, and $\pi^+ \pi^+$
pairs in collisions of 450 GeV p on Pb \cite{NA44SPb}.
\end{figure}

\begin{figure}
\begin{center}
\epsfig{file=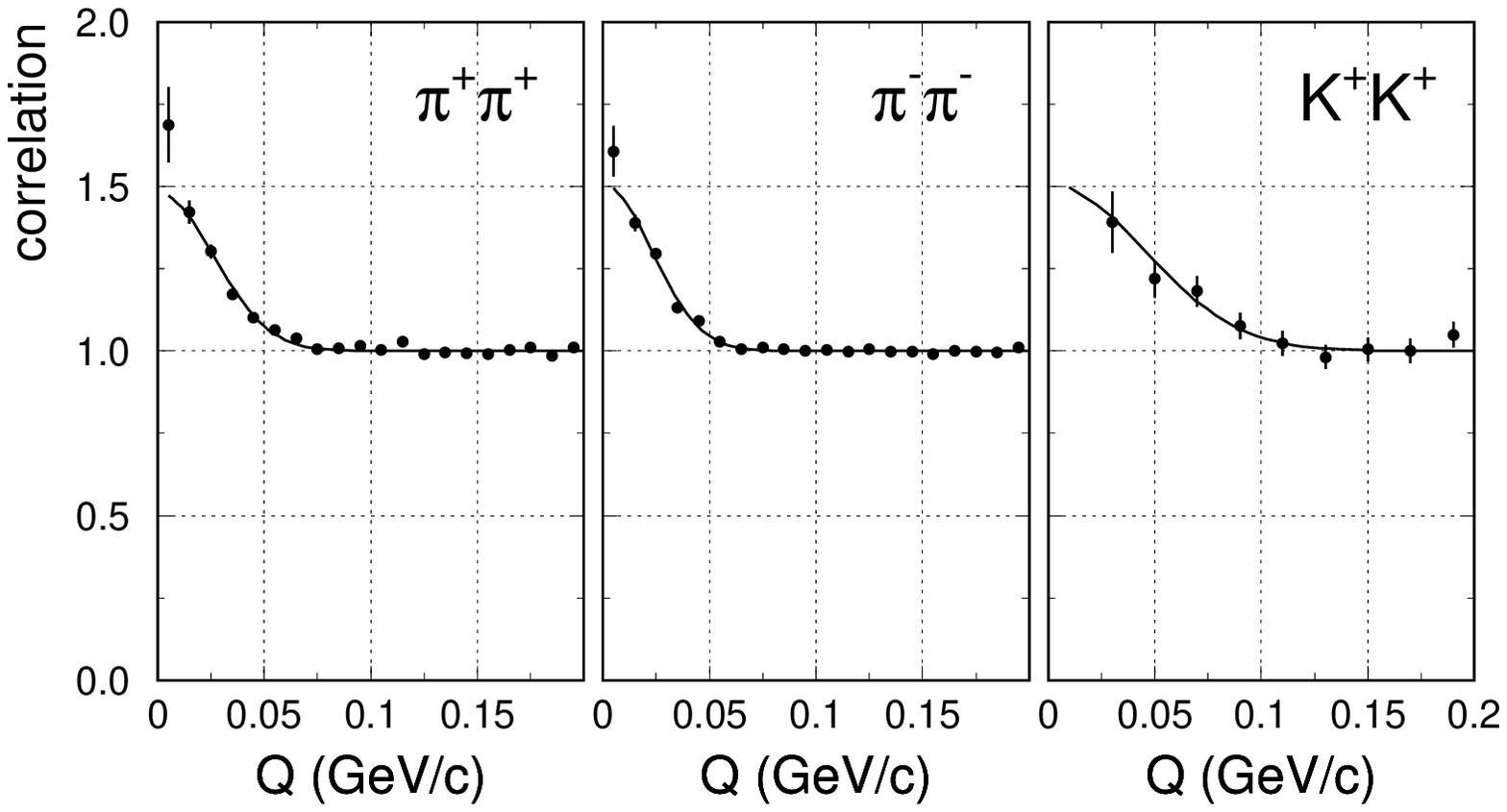, height=7cm}
\end{center}
Fig.~2. Two-particle correlation function for $\pi^+ \pi^+$, $\pi^-
\pi^-$, and K$^+$K$^+$ pairs in collisions of Au on Au at 10.8 GeV/A
\cite{e877c}.
\end{figure}

    I will begin by describing the HBT effect in the simplest model of
classical waves, and then discuss how one can understand HBT in terms of the
quantum mechanics of the particles reaching detectors.  Then I will turn to
the nuclear physics applications and finally mention applications of HBT
interferometry in both atomic and condensed matter physics.  My aim here is to
describe the physics underlying the HBT effect.  For more detailed discussions
of the current experimental situation in ultrarelativistic heavy-ion
collisions and its theoretical interpretation, the reader is referred to,
e.g., the reviews [3-8].

\section{Basic model of HBT intensity interferometry}

    HBT interferometry differs from ordinary amplitude interferometry in that
it does not compare amplitudes (as in a Young's two-slit interferometer) but
rather intensities at different points.\footnote{As we shall see below, there
is a close connection between the two types of interferometry.} The simplest
picture of HBT interferometry, from which we can see the fundamental idea, is
to consider two distant random point sources of light, $a$ and $b$ (of the
same frequency), or more realistically for a star, a distribution of point
sources, and imagine measuring the light falling in two independent telescopes
1 and 2; \cite{LQM} see Fig. 3. The detectors are not connected by any wires.
Assume that the sources are separated in space by $\vec R$, the two detectors
by $\vec d$, and that the distance from the sources to the detectors, $L$, is
much larger than $R$ or $d$.

    Imagine that source $a$ produces a spherical electromagnetic wave of
amplitude $\alpha e^{ik|\vec r-\vec r_a|+i\phi_a}/|\vec r-\vec r_a|$, and
source $b$ a spherical wave of amplitude $\beta e^{ik|\vec r-\vec
r_b|+i\phi_b}/|\vec r-\vec r_b|$, where $\phi_a$ and $\phi_b$ are random
phases (we ignore polarizations here).  Let us calculate the correlation of
the electromagnetic intensities in 1 and 2 as a function of the separation of
the two telescopes.  The total amplitude at detector 1 is
\beq
A_1 = \frac{1}{L} \left(
\alpha e^{ikr_{1a}+i\phi_a} + \beta e^{ikr_{1b}+i\phi_b}\right),
\label{ampl}
\eeq
where $r_{1a}$ is the distance from source $a$ to detector 1, etc., and the
total intensity in 1 is
\beq
I_1
%\hspace{330pt}\nonumber \\
= \frac{1}{L^2} \left(|\alpha|^2 + |\beta|^2 +
\alpha^*\beta e^{i(k(r_{1b}-r_{1a})+\phi_b-\phi_a)}
+\alpha\beta^* e^{-i(k(r_{1b}-r_{1a})+\phi_b-\phi_a)}\right),
\nonumber \\
\eeq
with a similar result for $I_2$.  On averaging over the random phases the
latter exponential terms average to zero, and we find the average intensities
in the two detectors,
\beq
\langle I_1\rangle = \langle I_2\rangle =
 \frac{1}{L^2} \left(\langle|\alpha|^2\rangle +
\langle|\beta|^2\rangle\right).
\eeq
The product of the averaged intensities $\langle I_1\rangle \langle
I_2\rangle$ is independent of the separation of the detectors.

\begin{figure}
\begin{center}
\epsfig{file=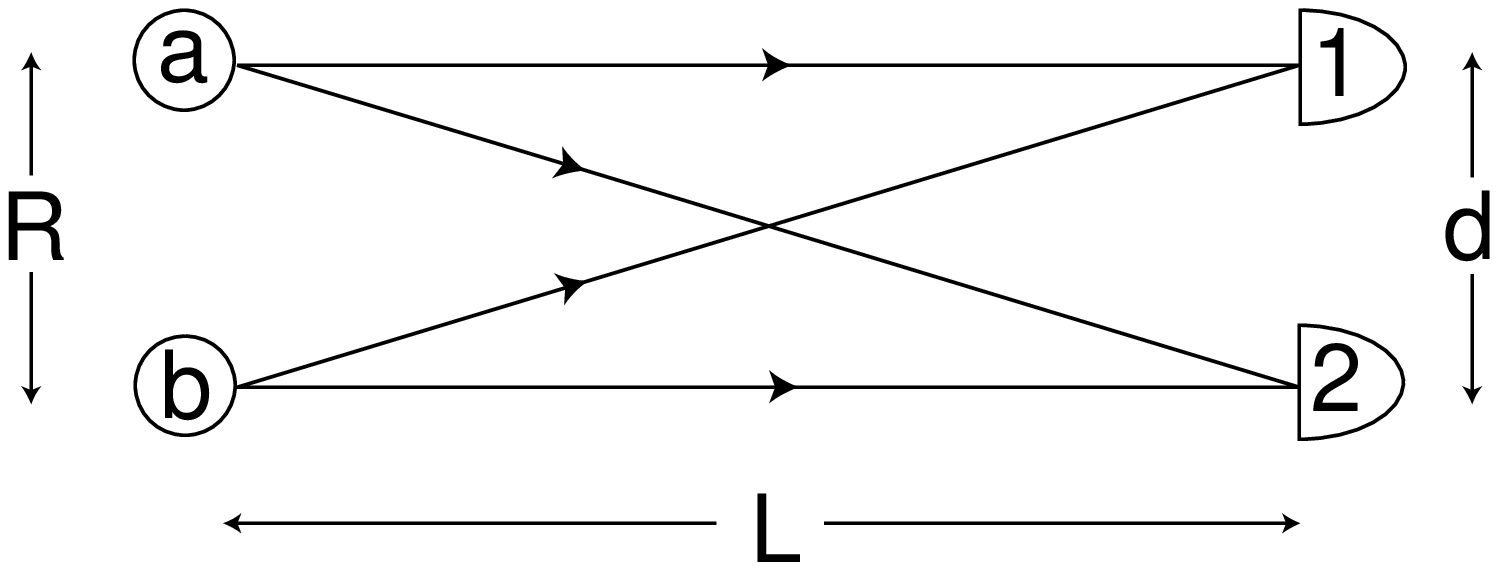, height=3cm}
\end{center}
    Fig.~3. Measurement of the separation of two sources, $a$ and $b$, by
correlation of intensities in detectors 1 and 2.
\end{figure}

    On the other hand, multiplication of the intensities $I_1 I_2$ before
averaging gives an extra non-vanishing term $\sim
(\alpha^*\beta)(\alpha\beta^*)$, and we find after averaging over the phases
that
\beq
\langle I_1 I_2\rangle =
\langle I_1\rangle \langle I_2\rangle
 + \frac{2}{L^4} |\alpha|^2 |\beta|^2
\cos\left(k (r_{1a}-r_{2a}- r_{1b}+r_{2b})\right) \nonumber \\
=  \frac{1}{L^4}\left[ (|\alpha|^4+|\beta|^4)
+ 2|\alpha|^2|\beta|^2
(1+ \cos\left(k (r_{1a}-r_{2a}- r_{1b}+r_{2b})\right)\right].
\label{I1I2}
\eeq
Then
\beq
C(\vec d\,\,)
= \frac{\langle I_1 I_2\rangle}{\langle I_1\rangle \langle I_2\rangle}
%\nonumber \\
= 1 +
2\frac{\langle|\alpha|^2\rangle\langle|\beta|^2\rangle}
{(\langle|\alpha|^2\rangle +\langle|\beta|^2\rangle)^2}
\cos\left(k (r_{1a}-r_{2a}- r_{1b}+r_{2b})\right).
\label{2sources}
\eeq

    For large separation between the sources and detectors ($L\gg R$),
$k(r_{1a}-r_{2a}- r_{1b}+r_{2b}) \to k(\vec r_a - \vec r_b)\cdot( {\hat r_2} -
{\hat r_1}) = \vec R\cdot (\vec k_2 -\vec k_1)$, where $\vec k_i= k{\hat r_i}$
is the wavevector of the light seen in detector $i$.  The correlated signal in
Eq.  (\ref{2sources}) varies as a function of the detector separation $d$ on a
characteristic length scale
\beq
 d = \lambda/\theta,
\eeq
where $\lambda$ is the wavelength of the light, and $\theta = R/L$ is the
angular size of the sources as seen from the detectors.  Thus by varying the
separation of the detectors one learns the apparent angle between the two
sources, and with a knowledge of the individual wavevectors, the physical size
of the source.

    If instead of two discrete sources, one has a distribution of sources,
$\rho(\vec r\,)$, then averaging over the distribution, one finds that the
correlation function measures the Fourier transform of the source
distribution:
\beq
C(\vec d\,\,)-1 \sim \left|\int d^3r \rho(\vec r\,)
e^{i(\vec k_1 -\vec k_2\,)\cdot\vec r}\,\right|^2.
\eeq

    One important difference between astronomical observations and high energy
physics is that the stars stay fixed, while in a collision, the system evolves
a time scale of 10$^{-23}$ to 10$^{-22}$ seconds, and thus one has to take
into account the changing geometry.  As we will see, in high energy physics
one measures not the Fourier transform of the distribution in space alone, but
to good approximation the Fourier transform in both space and time.  A second
important difference is that in astronomy, in the absence of a knowledge of
the distance to the source, one cannot measure the actual difference in
direction of the wavevectors of the light in the two detectors, and thus one
measures only the angular size of the source as seen from the detectors.  In a
high energy experiment, one can determine the wavevectors of the detected
particles, and thus measure the absolute size of the source.

    To find an enhanced correlation at detector separation $\le
\lambda/\theta$ it is not necessary for the two detectors to be wired
together.  One needs only to compare the data trains.  Why, we may ask, do
two independent nearby detectors produce extra signal?  Essentially if
the amplitude varies randomly then a positive fluctuation of the amplitude
will produce a correlated increase in both measured signals, and vice versa
for a negative fluctuation.  For example, in black-body radiation, both the
real and imaginary parts of the complex electric fields $E \sim \alpha
e^{i\vec k\cdot\vec r -i\omega t}$ are Gaussianly distributed.  For
independent Gaussianly distributed real variables $x$ and $y$, one finds
simply that $\langle (x^2+y^2)^2 \rangle = 2\langle x^2+y^2 \rangle^2$, so
that
\beq
\langle |E_1|^2 |E_1|^2 \rangle = 2(\langle |E_1|^2 \rangle)^2,
\label{efluct}
\eeq
while for a coherent source, e.g., a laser, $\langle |E_1|^2 |E_1|^2 \rangle
\simeq (\langle |E_1|^2 \rangle)^2$.  The extra factor of two is precisely the
source of the HBT correlations, the enhancement that arises from the cosine
term in Eqs.  (\ref{I1I2}) and (\ref{2sources}).

    Not apparent in the simple model above is how to deal with the time
involved in making measurements.  For example, how far apart can one shift the
data trains in time in comparing the intensities in the two detectors and
still find a correlation between the signals?  I will return to these
questions below.

\section{A brief history of the HBT effect}

    The radar technology developed in the Second World War opened the field of
radio astronomy in the postwar period, and soon led to the discovery of bright
radio ``stars" in the sky.  One had no idea how big various sources, e.g.,
Cassiopeia A and Cygnus A, were, and the immediate problem was how to measure
their sizes.  The standard technique in use was Michaelson interferometry, in
which one compares the {\it amplitudes} of the light landing at two separated
points, e.g., by converging the two signals using a lens and producing a
diffraction pattern as a function of the separation of the points.  From the
structure of the diffraction pattern (on a distance scale $\lambda/\theta$)
one can determine the angular size of the source.  Using this technique
Michaelson measured the angular diameter of Jupiter's system of moons in 1891,
and K. Schwarzschild first measured the angular diameter of binary stars in
1895.  The resolution by amplitude interferometry at a given wavelength is
limited by the size of the separations over which one can compare amplitudes.
Were the radio sources to have had a large angular size, then one would only
have needed a small separation of the two detectors.  On the other hand, were
the sources small, then it might have been necessary to separate the
telescopes by distances too large, e.g., on opposite sides of the Atlantic, to
be able correlate the amplitudes with the technology available in this period.
This is the problem that the radio astronomer Robert Hanbury Brown at Jodrell
Bank solved in 1949.  His basic realization was that ``if the radiation
received at two places is mutually coherent, then the fluctuation in the {\it
intensity} of the signals received at those two places is also correlated.''
\cite{boffin} Hanbury Brown then brought in Richard Twiss who had a more
mathematical training to carry out the mathematical analysis of intensity
correlations.

    The first test of intensity interferometry was in 1950, when Hanbury Brown
and Twiss measured the diameter of the sun using two radio telescopes
operating at 2.4m wavelength (in the FM band) -- quite a spectatacular
demonstration of the technique.  Their group then went on to measure the
angular diameters of the Cas A and Cyg A radio sources, which turned out to be
resolvable within a few kilometers.  Since they could in fact have done
Michaelson interferometry over such distances, Hanbury Brown described the
intensity interferometry effort as ``building a steam roller to crack a nut.''
\cite{boffin1} Nowadays, Michaelson interferometry has completely replaced
intensity interferometry in astronomy.  In radio astronomy, amplitude
interferometry is the basis of the Very Large Array (VLA) in Socorro, New
Mexico, and the extended VLBI, in which one compares radio amplitudes in
separated radio telescopes.  The 10 m optical Michaelson interferometer on the
Space Interferometry Mission satellite, to be flown in 2004, will be able to
resolve objects to 5 microseconds of arc.

    Intensity interferometry actually has an intimate relation with Michaelson
amplitude interferometry, as noted by Hanbury Brown and Twiss~\cite{HBT54}.
Amplitude interferometry measures essentially the square of the sum of the
amplitudes $A_1$ and $A_2$ falling on detectors 1 and 2:
\beq
|A_1+A_2|^2 = |A_1|^2 + |A_1|^2 + (A_1^*A_2 + A_1A_2^*).
\eeq
The latter term in parentheses, called the ``fringe visibility," $V$, is the
interesting part of the signal.   Averaged over random
variation in the signal, $V^2$ is simply
\beq
\langle V^2\rangle = 2\langle |A_1|^2 |A_2|^2 \rangle +
\langle A_1^{*2} A_2^2 \rangle + \langle A_1^2 A_2^{*2} \rangle.
\label{mich}
\eeq
As one can see from the simple model above, Eq. (\ref{ampl}), the final terms
vary rapidly on a scale of separations, $d\sim \lambda$, the wavelength
of the radiation, and average to zero.  On the other hand, the first term in
Eq. (\ref{mich}) is just twice the correlation of the intensities landing in
the two detectors.   Thus
\beq
\langle V^2\rangle \to 2\langle I_1 I_2\rangle;
\eeq
the time-average of the square of the fringe visibility is proportional
to the time-averaged correlation of the intensities.

    While it was well demonstrated both theoretically and experimentally that
intensity interferometry worked for radio waves, which were commonly
understood as classical fields, it was not obvious in the early 1950's that
the effect should also work for light.  Light being made of photons was more
mysterious than radio signals made of classical electrical waves; the
connections, now clear, were obscure at the time.  Hanbury Brown and Twiss
decided to test the idea for optics, with a simple tabletop experiment in
which they used a beam from a mercury vapor lamp -- a thermal source -- and a
half-silvered mirror to split the beam in two~\cite{bunch}.  By measuring the
intensity correlations between the two separated beams, they essentially
compared the intensities at two different points in the unseparated beam, and
by varying the relative path lengths between the mirror and the detectors they
could vary the time separation, $\tau$, of the points.  What they found was
that while at large $\tau$ there were no intensity correlations, the
correlations increased with decreasing $\tau$.  The characteristic timescale
is the {\it coherence time} of the beam which, in this case is essentially
$\hbar/T$, where $T$ is the temperature of the source.  This experiment was
the crucial demonstration of ``photon bunching," i.e., that photons in a
seemingly uncorrelated thermal beam tend to be detected in close-by pairs.
Their results were greeted with great disbelief, and various experiments were
done to disprove them.  In the end Hanbury Brown and Twiss prevailed, aided by
a particularly important paper by Purcell \cite{purcell} which showed how to
understand the effect in terms of electric field fluctuations (see Eq.
(\ref{efluct})) -- and the field of quantum optics was born.

    Armed with the demonstration that intensity interferometry worked for
light, Hanbury Brown and Twiss then went on to apply the technique to measure
the angular size of the star Sirius ($\alpha$ Canus Majoris A) by studying
optical intensity correlations between two telescopes \cite{sirius}.  Since
the telescopes required good light gathering ability but not great resolution,
Hanbury Brown and Twiss were able to fashion a pair from five-foot diameter
searchlights left over from the Second World War.  The signals from the two
telescopes were correlated electronically (although the actual physical
connection is not needed to observe the effect).  Figure 4 sketches their data
for the correlation function $C(d)-1$, divided by its value at $d$=0, measured
as a function of the separation $d$ of the two telescopes out to a maximum
separation $\sim$ 9 m. The data yielded an angular diameter of Sirius of
0.0068$^{\prime\prime} \pm$ 0.0005$^{\prime\prime} = 3.1\times 10^{-8}$
radians, a very impressive measurement of an object at a distance of 2.7 pc.
The four data points shown were taken for a total of some 18 hours over a 5
month period, an indication of the poor viewing conditions.  The dashed line
is the expected curve for a uniformly illuminated disk of angular diameter,
0.0063$^{\prime\prime}$.

\begin{figure}
\begin{center}
\epsfig{file=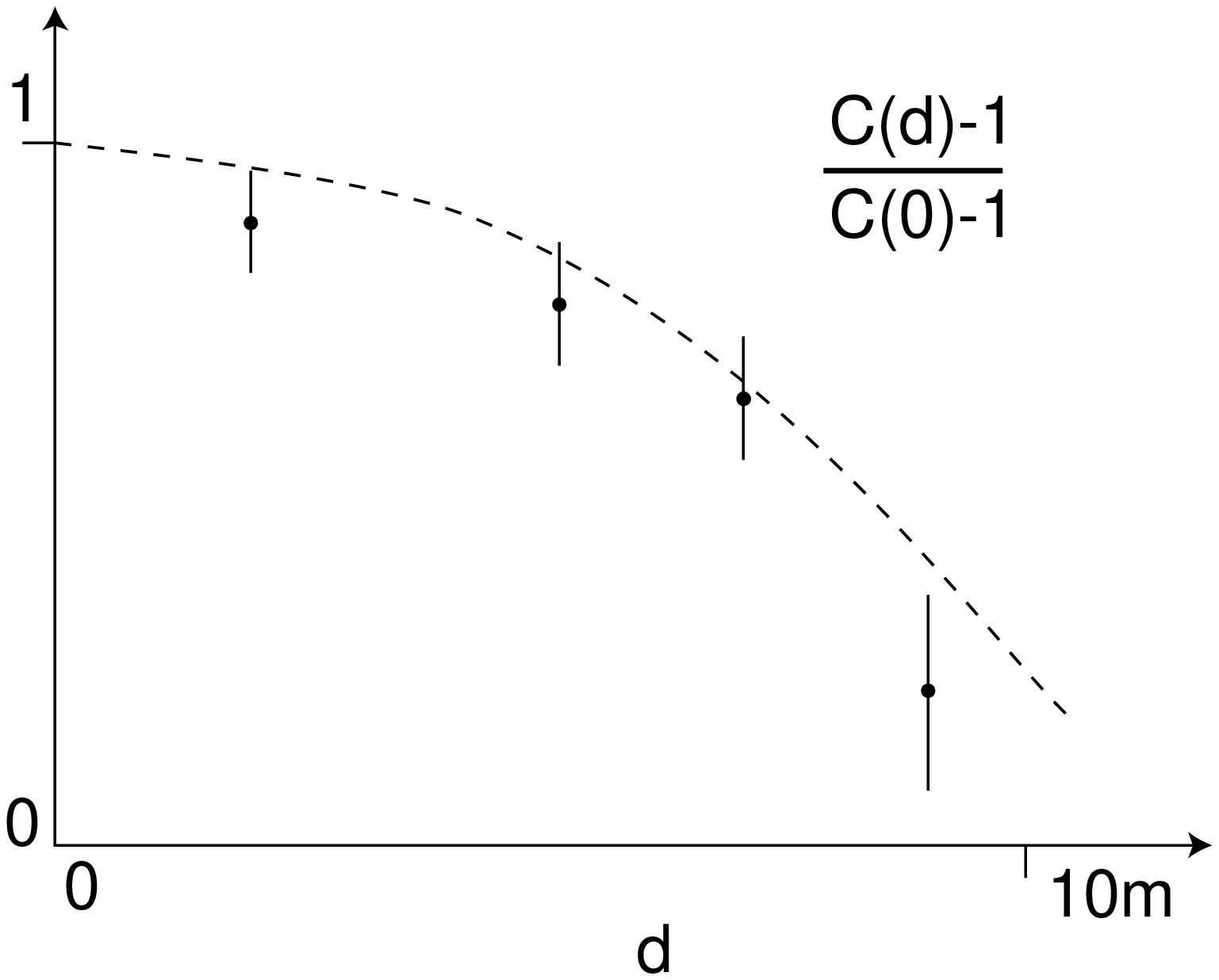, height=5.5cm}
\end{center}
Fig.~4.  Measurement of the angular diameter of Sirius \cite{sirius}.
\end{figure}

    This figure looks very similar in structure to the heavy-ion plots in
Figs. 1 and 2. An important difference is that the actual HBT correlation seen
here was just one part in $10^6$, a tiny signal above the background.  What is
the source of this difference?  The question is whether all observed pairs of
photons are ``HBT-correlated"\footnote{Here we mean correlated in the sense
that the photon pairs will produce an HBT effect at the detectors, measured by
$C$, as opposed to the different question of the correlations in the beam
produced in the source, as discussed in Sec. 6. The language is potentially
confusing, since photons that are maximally correlated at the source, e.g., in
a laser beam, do not exhibit an HBT effect, while a thermal source, which is
minimally correlated, produces the maximum HBT effect.}; for example, if one
takes a data train from 1956 in one telescope and compares it with a data
train in the other from 1997 will one see interferometry?  The answer is that
the photons are in fact HBT-correlated only if they are emitted within a
coherence time, the characteristic timescale in the original HBT tabletop
experiment with an optical source.  For a star the coherence time is
$\tau_{\rm coh} \sim 10^{-14}$ sec.  On the other hand, the signal was studied
over a band 5 - 45 MHz, corresponding to a binning time $\tau_{\rm bin} \sim
10^{-8}$ sec.  Roughly the probability of observing an HBT-correlated pair of
photons is $\sim \tau_{\rm coh}/\tau_{\rm bin}\sim 10^{-6}$.  Figure 5 shows
the region where photons produce an HBT signal in the plane of the times,
$t_1$ and $t_2$, of detections in detectors 1 and 2. Below we discuss the
analogous timescales in heavy-ion collisions.

\begin{figure}
\begin{center}
\epsfig{file=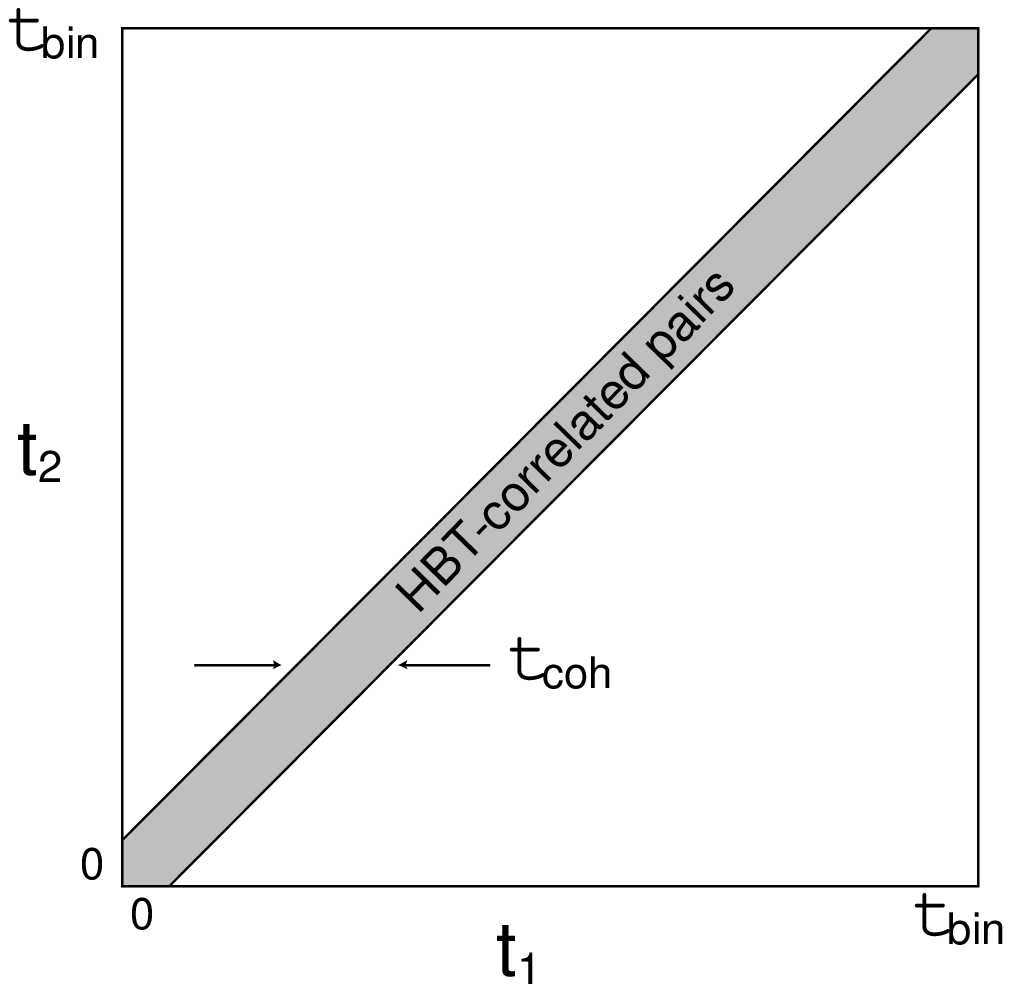, height=8cm}
\end{center}
Fig.~5. Region in the plane of the detection times $t_1-t_2$ where photon
pairs produce an HBT signal.
\end{figure}

\section{Quantum mechanics of HBT}

    The simple derivation of intensity interferometry in Sec. 2 is entirely
classical.  How can one understand the effect from a quantum mechanical
viewpoint?  If we think of the sources $a$ and $b$ in Fig. 3 as emitting
photons, we can identify four different processes, shown in Fig. 6:  i) source
$a$ emits two photons, one detected in each detector, ii) source $b$ emits
the two photons instead; iii) source $a$ emits a photon detected in 1 and $b$
emits a photon detected in 2, and finally, iv) source $a$ emits a photon
detected in 2 and $b$ emits a photon detected in 1, the exchange of the
previous process, iii.  The first two processes are distinguishable, and do
not produce any interferometry.  They simply correspond to detection of the
sources independently (the $|\alpha|^4$ and $|\beta|^4$ terms in Eq.
(\ref{I1I2})).  Only the latter two processes, iii and iv, which are
quantum-mechanically coherent, give rise to interferometry.  [Indeed, if we
drop the terms proportional to $|\alpha|^4$ and $|\beta|^4$, then Eq.
(\ref{2sources}) reduces to $ C(\vec d\,\,)= 1 + \cos(k (r_{1a}-r_{2a}-
r_{1b}+r_{2b}))$.] Quantum mechanically, the HBT effect is a
consequence of ordinary boson exchange, an effect included in the symmetry of
the wave function of the pair of particles, e.g., for a pair of independent
bosons in orthogonal states $\phi_\alpha$ and $\phi_\beta$, $\phi(1,2) =
\left(\phi_\alpha(1)\phi_\beta(2)+\phi_\alpha(2)\phi_\beta(1)\right)/\sqrt2$.
The effect is present for all pairs of identical bosons, including pions and
kaons produced in high energy collisions.

    The detection of interferometry in particle collisions dates from 1962
when G. Goldhaber, S. Goldhaber, W.~Y.  Lee, and A. Pais \cite{GGLP} studied
angular correlations of pions produced in p$\overline{\rm p}$ collisions at
the Bevatron.  According to Pais \cite{pais}, the idea of exploring
interferometry in particle physics, although so similar to that in
astronomical observations, was independently conceived.  The method is now a
standard technique in high energy collisions, from heavy ions [3-8], to
meson-nucleon interactions \cite{NA22}, to electron-positron annihilation
\cite{LEP}.  As noted by Feynman~\cite{feynman}, the experiment done with
electrons would yield intensity anti-correlations.  However the effect is
obscured by interactions among the fermions; electron pairs or proton pairs
have a repulsive Coulomb interaction which itself decreases the correlation
function at small momentum differences (see Sec. 9), while neutrons at small
relative momenta have significant final state strong interactions.
Correlation studies of nucleon pairs produced in heavy-ion collisions are
described in Refs.  \cite{bauer} and \cite{westerberg}, and references
therein.  HBT interferometry is now being applied in study of boson atomic
beams as well \cite{shimizu}, as we discuss in Sec. 10.

    Eventually, we will describe HBT measurements in terms of the two-particle
correlation functions of the emitted identical particles.  It is more
intuitive, however, first to study the problem in terms of particle wave
functions.  To be specific we focus on pions, although the discussion is quite
general.

\begin{figure}
\begin{center}
\epsfig{file=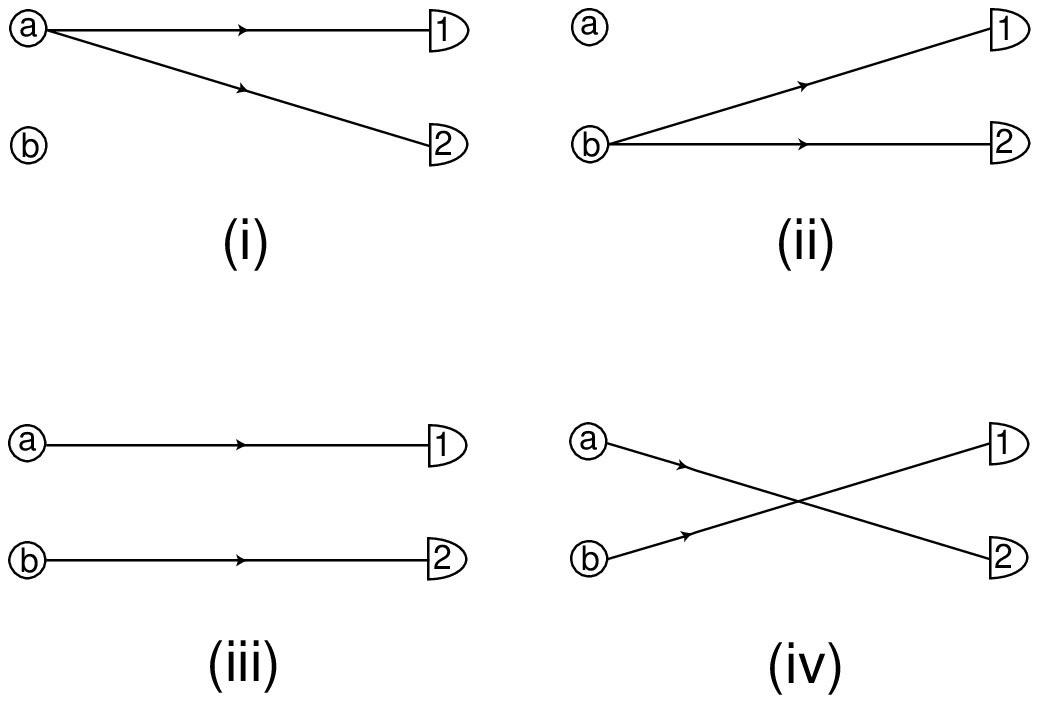, height=7cm}
\end{center}
Fig.~6.  The four independent photon emission and detection processes included
in Eq. (\ref{I1I2}).
\end{figure}

    How does one describe quantum mechanically the set of pions emitted
emitted in a nuclear collision?  Even in the best of all possible worlds --
where one knows the exact wave function of the two colliding nuclei, and knows
exactly how the quantum mechanical evolution operator does its job to produce
the final system as a coherent superposition of well defined pure quantum
mechanical states $\Psi(1,2,...,N)$ of $N$ particles -- the subset of pions
emitted is described by a {\it mixed} quantum state.  Quite generally, any
subset of particles in a pure state is described by a mixed state, even, e.g.,
for the electron in the ground state of a freely moving hydrogen atom.  The
single particle density matrix for pions of a given charge at equal time is
given by
\beq
\langle \pi^\dagger(r,t)\pi(r',t)\rangle =
\int d^3r_2 \cdots d^{3}r_n
\Psi^*(r,r_2,\cdots,r_N,t) \Psi(r',r_2,\cdots,r_N,t),
%\nonumber \\
\label{singden}
\eeq
where $\pi(x)$ is the operator destroying a pion of the given charge at
point $x = (\vec r\,,t)$.

    If $\Psi$ is a product of single particle wave functions, then
$\langle \pi^\dagger(x)\pi(x')\rangle$ factors into a product of single
particle states, $\phi^*(x) \phi(x')$.  In general the single-pion correlation
function does not factor, even when the pions are no longer interacting, but
it can be represented as a sum over a collection of single particle states
$\phi_i$:
\beq
\langle \pi^\dagger(x)\pi(x')\rangle =
\sum_i f_i \phi_i^*(x) \phi_i(x'),
\label{singden1}
\eeq
where the $f_i$ give the probability of finding the pion in single
particle state $i$.  For example, the probability of finding a pion at point 1
is given by $\sum_i f_i|\phi_i(1)|^2$.  Only if the pion part of the state
$\Psi$ factors out in the form of a product of the same single particle states
for all the pions -- a Bose-Einstein condensate -- will single pions be in a
pure state.  The mixed single pion state always has finite entropy,
$-\sum_i\left(f_i\ln f_i - (1+f_i)\ln (1+f_i)\right)$.

    The closest one can come to describing pions as little bullets is to
picture the single particle states making up the mixed ensemble of pions as
wavepackets with almost well defined momenta and energies, limited by the
uncertainty principle.  We picture the collision volume as made up of many
sources of pions; whether the sources are fragmentation of strings, or in the
language of low energy nuclear physics individual nucleon-nucleon collisions,
the sources are localized to within a distance $R$ which is less than the
size of the entire collision volume, and the emission process is temporally
localized to within a time $\tau$.  Thus the individual components of
momentum, $\vec p\,$, and energy, $\eps_p$, of the emitted particles are
uncertain to within
\beq
\Delta p_a &\ga& \hbar/R, \quad a= x,y,z \nonumber \\
\Delta \eps_p &\ga& \hbar /\tau.
\label{uncertain}
\eeq

    A pion nominally of momentum $\vec p\,$ emitted from a source at the
origin in space and time would have an amplitude to have four-momentum $q$ =
($\eps_q$, $\vec q\,$) that is roughly Gaussian, \beq
 A(q) \sim e^{-(\vec q - \vec p\,)^2R^2/2}
e^{-(\eps_q - \eps_p)^2\tau^2/2},
\label{gaussian}
\eeq
and in space-time the particle would be described by a wavepacket
\beq
\phi_{\vec p}\,(x) = \int \frac{d^3q}{(2\pi)^3} \frac{e^{iqx}}{2\eps_q} A(q).
\label{packet}
\eeq

    How does the packet evolve in time after leaving the source?  Assume that
the collision takes place, and the particles emerge, into vacuum.  (In real
life secondary scattering on other atoms in the target and also scattering in
air are both very important effects, to which we return in Sec. 8.)  The
transverse spatial spread, perpendicular to $\vec p\,$, is determined by the
uncertainty in transverse velocity, which for for relativistic particles
($\eps_p\approx p$) is:
\beq
\Delta v_\perp = \frac{\Delta p_\perp}{\eps_p} \sim \frac{1}{pR}.
\label{vtrans}
\eeq
The spread in longitudinal velocity is
\beq
\Delta v_L = \Delta \left(\frac{p_L}{\eps_p}\right) = \frac{\Delta
p_L}{\gamma^2\eps p}
\sim \frac{1}{\gamma^2pR},
\label{vlong}
\eeq
where $\gamma = \eps_p/m$ is the Lorentz factor, and $m$ is the pion mass.
The presence of the factor $\gamma^2$ reflects the fact that the more
relativistic the particles become the closer to the speed of light the packet
moves, with vanishing spread in longitudinal velocity.  The transverse size of
the wavepacket after travelling a distance $L$ is thus $\sim L/pR$, while its
thickness is $\sim L/\gamma^2pR$.

    To be specific, consider a 1 GeV pion ($\gamma = 7$) produced within an
initial radius of 10 fm travelling to a detector 10 meters from the source.
Then the transverse spread of the wavepacket is about 20 cm, while the
thickness of the packet grows to half a centimeter.  Pions emerge in rather
extended pancake-like states.  The characteristic time for such a pion
wavepacket to cross a point at the distance of the detector is $\sim
10^{-11}$ sec.  Smaller source sizes lead to even larger spreads in the pion
wavepackets.  Photons would have a similar transverse spread; however, the
thickness of the wavepacket would not grow.

\section{Detector response}

    Let us now turn to the question of the detection of particles in such
wavepacket states by a magnetic spectrometer.  In the ``bullet" picture, the
particle travels along a classical trajectory, excites an atom in the
detector, which determines the direction of its momentum, and begins to make a
track (or in emulsion makes a spot on photographic film); the particle
continues on, producing a curved track in the spectrometer, from which one
deduces the magnitude of its momentum.  But, in fact, the particle has a
distribution of probability amplitudes, wherever its wavepacket $\phi$ is
non-zero, of where it makes the first spot in the spectrometer.  A particle of
mean momentum $\vec p$ can, because of the transverse momentum uncertainty,
be detected anywhere within the range of momenta $\Delta p_\perp$ around
$\vec p\,$ contained in its wavepacket.  After the initial collision in the
detector, a much more narrowly focused wave packet emerges.  (If one does not
actually do the measurement, the state that emerges is a mixed state
corresponding to all possible points where the incident packet can excite a
detector atom.)  The narrowed packet continues through the magnetic field, and
the subsequent collisions it makes selects its energy.  One measures momentum
by measuring a sequence of positions in the detection.

    Consider the measurement of the momentum of an incident particle in state
$\phi_i(x)$.  The probability that it makes the first single excitation of an
atom in the detector at point $A$ and continues to produce a track
corresponding to measuring its momentum to be $\vec k$ is then
\beq
P_{\vec k\,}(A;i) =
  \int dx dx' e^{ikx}\phi_i^*(x) s_A(x,x') e^{-ikx'}\phi_i(x').
\eeq
Here the detector atom response function, $s$, is localized in $r$ and
$r'$ about the atom at $A$; it also extends over an interval in $t-t'$ which
is the characteristic time that the detector analyzes the amplitude and phase
variation of the incident wave, of order $10^{-16}$ sec $\sim \hbar$/(10 eV),
the inverse of a characteristic atomic excitation energy.  This time is much
shorter than the typical nanosecond scale resolution time of a detector, the
time it takes to build up a million-fold cascade of electrons.\footnote{A
detailed discussion of the role of the detectors is given in J. Popp's thesis
\cite{popp}, and in Ref.~\cite{hbtabp}.} More generally, the probability of
detection of a pion of given charge at $A$ is given by \beq P_{\vec k\,}(A) =
\int dx dx' e^{ik(x-x')}s_A(x,x') \langle \pi^\dagger(x) \pi(x')\rangle, \eeq
where $\langle \pi^\dagger(x) \pi(x')\rangle$ is the single-pion correlation
function.

    Consider next the detection of two independent particles.  The question
is why, if one detector lights up, do nearby detectors tend to have greater
probability to light up than detectors further away -- the HBT effect.
Suppose first that the two particles are incident in orthogonal wavepackets
$\phi_i$ and $\phi_j$.  The symmetrized two-particle wave function is
$\phi(\vec r\,,\vec r\,',t) = \left(\phi_i(\vec r\,,t) \phi_j (\vec r\,',t)+
\phi_i(\vec r\,',t) \phi_j(\vec r\,,t)\right) /\sqrt2$.  The probability of a
joint detection by a detector atom at $A$ of a pion that continues to make a
track corresponding to momentum $\vec k$, and by a detector atom at $B$ of a
second pion that continues to make a track corresponding to momentum $\vec
k\,'$, is then given by
\beq
\lefteqn{P_{\vec k\,,\vec k'\,}(A,B;i,j)}
%\nonumber \\
&=& \int dx dx''
 e^{ik(x-x'')}s_A(x,x'') \int dx' dx'''  e^{ik'(x'-x''')}s_B(x',x''')
 %\times
\nonumber \\
   && \hspace{20pt}\times\left(\phi_i(x) \phi_j(x') + \phi_i(x')
  \phi_j(x)\right)^*
   \left(\phi_i(x'') \phi_j(x''') + \phi_i(x''') \phi_j(x'')\right)
\nonumber \\
   &=& P_{\vec k\,}(A;i) P_{\vec k\,'}(B;j)
+ P_{\vec k\,}(A;j) P_{\vec k\,'}(B;i)
\nonumber \\
   &&\mbox{} +
{\cal{A}}_{\vec k\,}(A;i,j){\cal{A}}_{\vec k\,'}(B;j,i) +
   {\cal{A}}_{\vec k\,}(A;j,i){\cal{A}}_{\vec k\,'}(B;i,j),
\label{2part}
\eeq
where
\beq
{\cal{A}}_{\vec k\,}(C;i,j)=
\int dx dx'' e^{ik(x-x'')}s_C(x,x'') \phi_i^*(x)\phi_j(x'').
\label{overlap}
\eeq
The first term in Eq.  (\ref{2part}) is the probability of the particle in
state $i$ being detected at $A$ times the probability of the particle in state
$j$ being detected at $B$, and the second is the same with $i$ and $j$
interchanged.  These are the normal terms.

    The final two terms in Eq.~(\ref{2part}) are the enhancement of the
detection probability -- the HBT effect.  As we see, in order to have
enhancement, it is necessary that both wavepackets, $i$ and $j$, overlap in
the detector at $A$ during the time that the detector is doing quantum
mechanics on the incoming system, and similarly that they also must overlap in
the detector at $B$ (but note that the wavepackets do not have to be present
in both detectors simultaneously).  The presence of the wavepackets
simultaneously in each of the detectors is the reason the detectors produce
more signal when they are close to each other.\footnote{Imagine that the
detection at $A$ occurs before the detection at $B$.  Then one may ask how
both original wavepackets can be at the second detector, since the first
detection ``reduces" the wavepacket of the detected particle.  From this point
of view the amplitude for detection of a particle at $B$ is proportional to
the amplitude, $\phi_j(B)$, for it to be in state $j$ at $B$ times the
amplitude, $\phi_i(A)$, for the other particle in state $i$ to have been
detected earlier at $A$, plus the same product with $i \leftrightarrow j$.
The resulting joint probability is the same as Eq.  (\ref{2part}).  I thank J.
Walcher for raising this question.}

    The maximum transverse separation $d$ of the detectors that will produce
an HBT signal is essentially the transverse size of a given wavepacket,
$L/pR$, the wavelength divided by the angular size of the source as seen from
the detectors, where again $L$ is the distance from the source to the
detector, $R$ is the size of the source, and $p$ is the average particle
momentum.  The scale of relative momenta $q$ for which one finds a signal is
$q/p = d/L \sim 1/pR$, and thus $q \sim 1/R$, the standard HBT result.  As we
see from this argument, the HBT effect directly measures the width of the
wavepackets at the first detection.  This width is in turn determined by the
uncertainties in the momentum distribution at the time that the wavepacket
is no longer affected by strong interactions with the other particles in the
collision..

    Generally the amplitudes, Eqs.  (\ref{gaussian}) and (\ref{packet}), of
the wavepackets, $A(q)$, vary slowly in $q$ on an atomic scale, $\sim$ 1
KeV/c.  In this case the overlap integral (\ref{overlap}) is not sensitive to
the detection time scales, $\sim 10^{-16}$ sec, and one finds a correlation
function:
\beq
C(p,p') =  1 + \frac{\left|\sum_i f_i A_i^*(p) A_i^*(p')\right|^2}
                      {\sum_i f_i |A_i(p)|^2 \sum_j f_j |A_j(p')|^2},
\label{cf}
\eeq
where $f_i$ is the single pion probability in the ensemble, Eq.
(\ref{singden1}).

    However, if a particle is delayed in emission by more than the detection
time scale, the HBT correlations between that particle and one produced
directly will be suppressed.  A simple example in heavy-ion collisions of this
effect is in the correlation of $\pi^-$ produced in $\Lambda$ decay, $\Lambda
\to \pi^- + p$, with directly produced pions.  Because the $\Lambda$ travels
more slowly than a directly produced pion of the same rapidity as one emitted
in the decay, the pion from decay will lag the directly produced one by a time
$\Delta t$.  To estimate this effect, we note that a $\pi^-$ emitted in the
forward direction has rapidity $y_\pi^0 \approx 0.67$ in the $\Lambda$ frame,
and that a $\Lambda$ of rapidity $y$ travels on average a distance
$\tau_\Lambda \sinh y$ before decaying, where $\tau_\Lambda$ is the $\Lambda$
lifetime.  Thus $\Delta t = \tau_\Lambda/(\cosh y + \sinh y/ \tanh
y_{\pi}^0)$, which for a $\Lambda$ of typical rapidity 3 is $\sim 0.037
\tau_\Lambda = 9.7 \times 10^{-12}$ sec, much longer than the detector
timescale.  Pions emitted in other than the forward direction will have an
even greater time lag.

\begin{figure}
\begin{center}
\epsfig{file=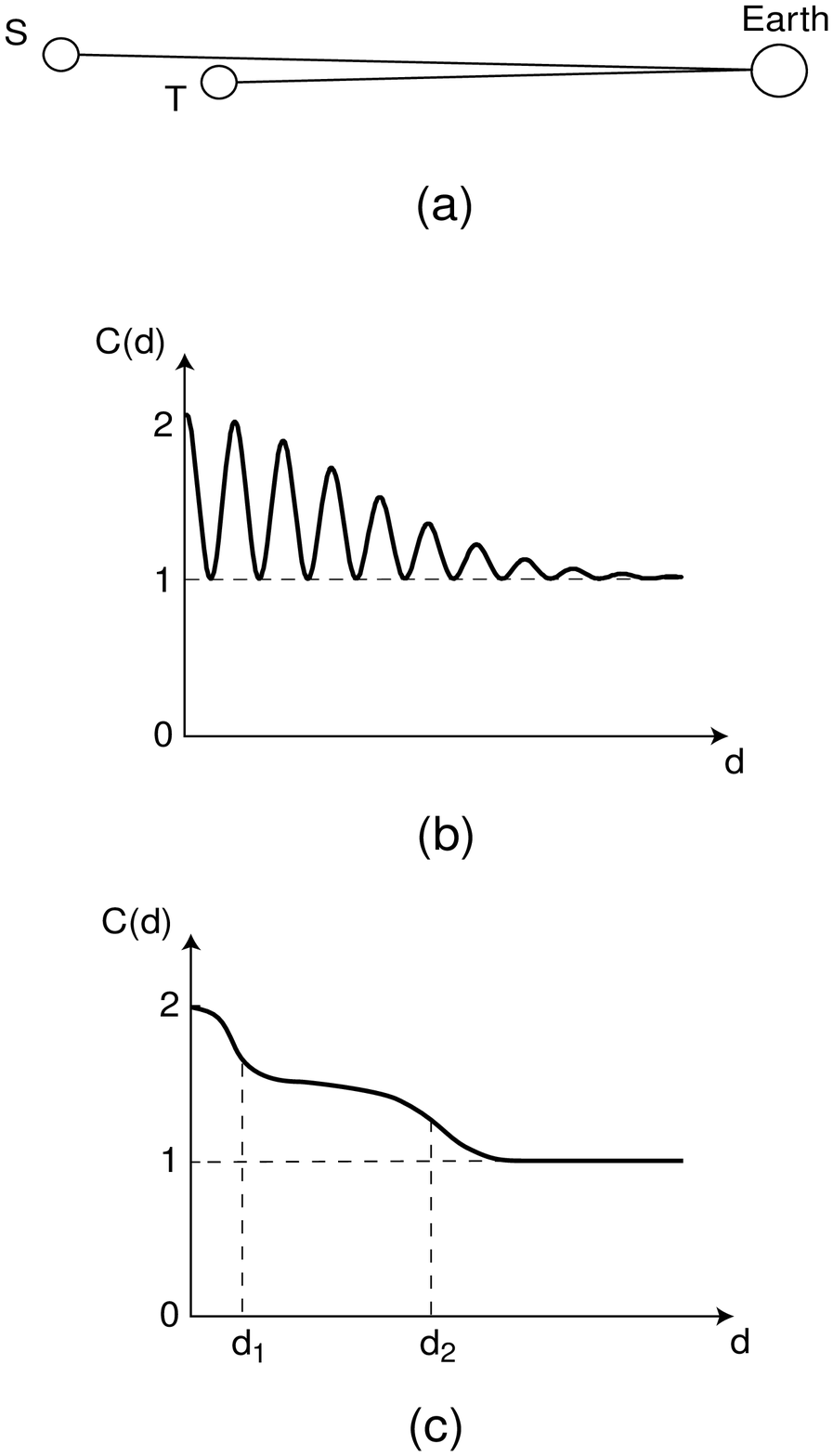, height=10cm}
\end{center}
Fig.~7. a) Two stars, $S$, and $T$, along nearby lines of sight from the
earth; b) schematic of HBT measurement of correlated intensity from the two
stars; c) schematic of HBT measurement a bright star surrounded by a halo of
dim stars.
\end{figure}

\begin{figure}
\begin{center}
\epsfig{file=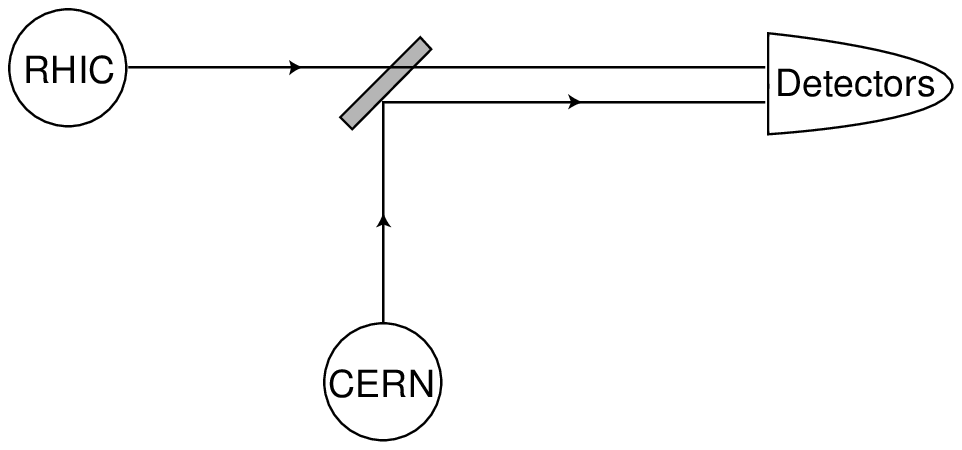, height=4cm}
\end{center}
Fig.~8.  Detection of the HBT effect between a pion from RHIC and a pion from
CERN.
\end{figure}

    As the above discussion makes apparent, the enhanced signal is not a
consequence of special preparation at the source, such as a particle of a
given momentum inducing emission of other particles of similar momentum, as in
a laser.  (In fact, a coherent source such as a laser would not give an HBT
signal.)  Clearly, in the stellar case there can be no such connection between
emission processes on opposite sides of the star, and yet photons from
opposite sides give an HBT enhancement.  The effect is a property of the
detection.  Furthermore, if two stars, $S$ and $T$, which are at very
different distances from the Earth, are approximately along the same line of
sight, Fig. 7a, one will see HBT correlations between a photon from $S$ and a
photon from $T$.  Of course the emission times have to be different in order
that the two photons arrive at each of the detectors at the same time.  (Even
in measuring a single star one correlates earlier emitted photons from the
stellar rim with later emitted photons from the front surface.)  The HBT
correlations between photons from the two sources as a function of detector
separation, $d$, sketched in Fig. 7b, would have an oscillatory term
characteristic of the angular separation of the two stars (cf.  Eq.
(\ref{2sources}), modulated by a broader Gaussian characteristic of the
angular size of an individual star.  By comparison, a single bright star, $T$,
surrounded by a halo of dim stars (an average over a distribution of stars
$S$), would yield a correlation function, Fig. 7c, with a rise proportional,
for $d\,\la \,d_2$, to the inverse of the angular size of the central star
$\lambda/d_2$, plus a much more narrow rise, for $d\,\la\, d_1$, inversely
proportional to the angular size of the halo.  As an exercise, sketch the
correlation function produced by two lasers which are mutually incoherent,
rather than the two stars.

    That the HBT effect is a not a property of production, but rather is a
property of detection is well illustrated by the following experiment.
Imagine, as in Fig. 8, that RHIC, the BNL heavy-ion collider, sends a pion
into a far away particle detector array, and that by using the pionic analog
of a half-silvered mirror, we reflect a pion from CERN into the same detector
array, so that it arrives at the same time as the pion from RHIC.  As long as
the wavepackets of the pions overlap sufficiently that the interference
amplitudes ${\cal{A}}$ at the detector atoms are non-zero, one will see an HBT
enhancement!  As this experiment illustrates, the HBT effect does not depend
on the history of the particles.  What matters is the form of the wavepackets
when they arrive at the detectors.

\section{Correlation functions}

    Let us now turn to describing the HBT effect in terms of the correlations
of the sources of pions in a collision.  To generalize Eq.  (\ref{2part}), we
may write the joint probability of detection at the detectors at $A$ and $B$
in terms of the two-pion correlation function $\langle \pi^\dagger(x)
\pi^\dagger(x'') \pi(x''')\pi(x')\rangle$ as
\beq
\lefteqn{P_{\vec k\,,\vec k'\,}(A,B)}
\nonumber \\
&=& \int dx dx'
 e^{ik(x-x')}s_A(x,x') \int dx'' dx'''  e^{ik'(x''-x''')}s_B(x'',x''')
\nonumber \\
   && \hspace{20pt}\times
  \langle \pi^\dagger(x) \pi^\dagger(x'') \pi(x''')\pi(x')\rangle.
\label{2partcorr}
\eeq
The two-pion correlation function $\langle \pi^\dagger(1) \pi^\dagger(2)
\pi(3)\pi(4)\rangle$, where the integers stand for space-time points, is the
amplitude for starting in a state of the system, removing a pion at 4, then
removing a second pion at 3, adding a pion back at 2, adding another back at
1, and returning to the initial state.

    Consider first a system of $N$ free bosons that is completely
Bose-Einstein condensed, or the photons produced by a laser.  For such a
system all the particles, or photons, are in the same particle state
$\phi(x)$.  Then the correlation function completely factors.  The amplitude
for removing a particle at a given point 4 is simply proportional to the wave
function $\phi$ at the point, while the amplitude for adding a particle is
proportional to $\phi^*$ at the point.  In this case the single-particle
correlation function is
\beq
\langle \pi^\dagger(1)\pi(2)\rangle = N\phi^*(1)\phi(2)
\eeq
and
\beq
\langle \pi^\dagger(1) \pi^\dagger(2) \pi(3)\pi(4)\rangle
&=& N^2\phi^*(1)\phi^*(2)\phi(3)\phi(4) \nonumber \\
&=& \langle \pi^\dagger(1)\pi(4)\rangle \langle \pi^\dagger(2)\pi(3)\rangle.
\label{coh}
\eeq
Such a source is {\it coherent}.\footnote{Taking Poisson statistics for
the distribution of the number $n$ of photons in a laser beam more carefully
into account leads to the same result (\ref{coh}), since $\langle
n(n-1)\rangle = \langle n\rangle^2$.}

    The correlations of particles from a thermal source, e.g., a black-body
cavity, are quite different.  For free particles, the mean number of particles
of energy $\eps$ is given by $\langle n_\eps\rangle =
1/(e^{\beta(\eps-\mu)}-1)$, while the fluctuations are given by
\beq
\langle n_\eps^2\rangle - \langle n_\eps\rangle^2 =
  \langle n_\eps\rangle(1\pm \langle n_\eps\rangle),
\label{thermfac}
\eeq
where the upper sign is for bosons and the lower for fermions.  (The
fermion result follows trivially since $n$ can only equal 0 or 1.) Translated
back into a statement about the correlation function, one finds
the expected factorized form for a thermal ensemble:
\beq
\langle \pi^\dagger(1) \pi^\dagger(2) \pi(3)\pi(4)\rangle
= \langle \pi^\dagger(1)\pi(4)\rangle \langle \pi^\dagger(2)\pi(3)\rangle
\pm \langle \pi^\dagger(1)\pi(3)\rangle \langle \pi^\dagger(2)\pi(4)\rangle,
%\nonumber \\
\label{factor}
\eeq
where $\pi$ here represents either a Bose or Fermi field.  This equation
says that if one removes two particles at 4 and 3, one can come back to the
same state by either replacing the first with a particle at 1 and the second
at 2 (first term), or the first at 2 and the second at 1 (second term).  In
the boson case, when 3=4 and 1=2, the amplitude for removing two particles is
just $2\langle \pi^\dagger(1)\pi(4)\rangle^2$.  The extra fluctuations -- the
factor of 2 here, or more generally the second term on the right side of
Eq.~(\ref{factor}) -- are the source of the HBT interferometry effect.  The
maximum HBT effect occurs when the correlation function factorizes in this
fashion.  Then one describes the source as {\it chaotic}.

    The source need not be thermal to factor this way.  The result
(\ref{factor}) always holds for non-interacting fermions, while for bosons it
is sufficient that no single particle mode $i$ be macroscopically occupied,
i.e., that all $\langle n_i \rangle$ are $\ll 1$.  The basic reason one
expects the correlation function in heavy-ion collisions to factorize as in
Eq.~(\ref{factor}) is that the pions undergo considerable rescattering in the
hot environment of the collision volume; the key is the destruction by
rescattering of phase correlations among the pions from the production
process.\footnote{A simple example that violates Eq.~(\ref{factor}) are the
correlations among pions radiated by a weakly interacting gas of nucleons.}

    It is useful to relate the pion correlation functions to correlations of
the sources of the pion field.  The freely propagating field $\pi(x)$ measured
at the detector is produced according to the field equation
\beq
(\Box^2 - m_\pi^2)\pi(x) = J(x),
\eeq
where $J(x)$ is the source of the field at the last collision.  Then
\beq
\pi(x) = \int dx' D(x,x') J(x') =\int \frac{d^3k}{(2\pi)^3}
\frac{e^{ikx}}{2i\eps_k} \int dx' e^{-ikx'} J(x'),
\eeq
where $D$ is the free pion Green's function, and the latter form holds
in the far field.  The single pion correlation function is related to the
source correlation function by
\beq
\langle\pi^\dagger(x)\pi(x')\rangle =
\int dx''dx'''D^*(x,x'')D(x',x''')\langle J^\dagger(x'')J(x''')\rangle.
\eeq
Generally $\langle\pi^\dagger(x)\pi(x')\rangle$ can be written in a
factorized form as a sum of wavepackets as in Eq.~(\ref{singden1}).

    The measured singles distribution of pions is then given in terms of the
source correlation function by
\beq
\eps_p\frac{d^3n}{d^3p} = \frac{d^3n}{d^2p_\perp dy}
= \frac{1}{2} \int dx dx' e^{ip(x-x')}\langle J^\dagger (x) J(x')\rangle,
\label{singles}
\eeq
and the measured pair distribution by
\beq
\eps_p\eps_p'\frac{d^6n}{d^3pd^3p'}
%\hspace{260pt} \nonumber \\ \hspace{30pt}
= \frac{1}{4} \int dx dx' e^{ip(x-x')} dx'' dx'''
e^{ip'(x''-x''')}
\langle J^\dagger (x) J^\dagger(x'')J (x''') J(x')\rangle.
\nonumber \\
\label{d6njjjj}
\eeq

    In the following let us assume a chaotic source, so that
\beq
\langle J^\dagger (x) J^\dagger(x'')J (x''') J(x')\rangle\hspace{90pt}
\nonumber  \\
=\langle J^\dagger (x) J(x')\rangle \langle J^\dagger (x'') J(x''')\rangle
+\langle J^\dagger (x) J(x''')\rangle \langle J^\dagger (x'') J(x')\rangle;
\eeq
then the pair distribution function becomes
\beq
\eps_p\eps_p'\frac{d^6n}{d^3pd^3p'}
= \frac{1}{4} \int dx dx'dx'' dx'''  e^{ip(x-x')} e^{ip(x''-x''')}
%\nonumber \\ \hspace{2cm} \times
\langle J^\dagger (x) J(x')\rangle \langle J^\dagger(x'')J (x''')\rangle
\nonumber \\
\hspace{2cm}  \times\left[ e^{ip(x-x')} e^{ip'(x''-x''')}
  + e^{ip(x-x''')}e^{ip'(x''-x')}\right].
\label{d6njj}
\eeq
We see that the HBT correlation function, defined by
\beq
C(q) =  \frac{d^6n/d^3pd^3p'}{(d^3n/d^3p)(d^3n/d^3p')},
\eeq
where ${\vec q}={\vec p}-{\vec p\,'}$ (and with implied separate averages
over the center-of-mass coordinates of the numerator and denominator, cf.
Eq.~(\ref{cq})), measures the structure of the current-current correlation
function.  Note that the information it provides is on the nature of
the source of particles after the last strong interactions, when the particles
begin to stream freely towards the detectors.

    The correlation function $\langle J^\dagger(x)J(x')\rangle$ contains two
length, and time, scales.  The range of the center-of-mass variables, $X=
(\vec r+ \vec r\,')/2, (t+t')/2$, are on the order of the size of the
collision volume, $R \sim$ 10 fm, and the duration of the collision, $\tau$,
also on the order of 5 - 10 fm/c.  On the other hand, the dependences in
$x-x'$ measure the space-time extent of the region in which the phase at a
point $x'$ is coherent with the phase of the current at $x$, a region of size,
$\xi_c$ and $\tau_c$ in space and time.  The lengths $\xi_c$ and
$\tau_c$, which determine the falloff of the singles distribution,
Eq.~(\ref{singles}), are typically on the order of one fm in space and one
fm/c in time, much shorter than the range in the center-of-mass variables.

    Such behavior is illustrated by the factorized form for $\langle,
J^\dagger(x)J(x')\rangle$,
\beq
\langle J^\dagger(x)J(x')\rangle =
e^{-(\vec r\,+\vec r\,')^2/8R^2} e^{-(t+t')^2]/8\tau^2}g(x-x'),
\label{JJfact}
\eeq
The singles distribution, from Eq.  (\ref{singles}), is then
\beq
\eps_p\frac{d^3n}{d^3p} \sim \int dx e^{ipx}g(x),
\label{single1}
\eeq
and the two-particle correlation function, assuming a chaotic source, is
\beq
C(q)= 1+ e^{-{\vec q\,}^2R^2}e^{-{q^0}^2\tau^2}
\frac{(d^3n/d^3K)^2}{(d^3n/d^3(K+q/2))(d^3n/d^3(K-q/2))},
\label{doubles1}
\eeq
where $K=(p+p')/2$.  As we see from this equation, the length measured in
HBT is modified from the length, $R\,$, governing the center-of-mass behavior
of
the current-current correlation function, $\langle J^\dagger(x)J(x')\rangle$,
due to the final factor, the ratios of the singles distributions.

    A particularly simple and illustrative model is the following.  Assume
that the particle production is described by a distribution of sources of size
$R_s$, $\tau_s$, at space-time points $x_s$, each producing pions in
wavepackets of mean momentum $\vec p\,$,
\beq
\phi_p(x-x_s)=\int \frac{d^3k}{(2\pi)^3}
\frac{e^{ik(x-x_s)}}{2\eps_k}e^{-(p-k)^2R_s^2/2},
\eeq
with probability $f(\vec p\,)$.  The spread in momenta in the individual
states $\phi_p(x)$ is of order $\hbar/R_s$.  If the sources are Gaussianly
distributed in space and time over a region of size $R_0, \tau_0$,
then from Eq.~(\ref{singden1}) the pion correlation function is,
\beq
\langle \pi^\dagger(x)\pi(x')\rangle
%\hspace{240pt}\nonumber \\
\sim \int \frac{d^3p}{(2\pi)^3} f(p)
 \int d^4x_s  e^{-r_s^2/2R_0^2} e^{-t_s^2/2\tau_0^2}
 \phi_p^*(x-x_s)\phi_p(x'-x_s).
\label{pack1}
\eeq
Carrying out the $x_s$ integrals explicitly, one readily finds that this model
is the same as that with a source of the form (\ref{JJfact}), where
\beq
R^2 = R_0^2 + R_s^2/2,\,\, \tau^2 = \tau_s^2+\tau_0^2/2,
\eeq
and
\beq
g(x) =  \int\frac{d^3p}{(2\pi)^3} f(p)e^{-ipx}e^{-x^2/4R_0^2}.
\eeq
The details of the individual wavepackets are all subsumed in the
current-current correlation function.  As this model illustrates, the length
scale describing the center-of-mass of the current correlation function is the
size of the distribution of sources, plus a correction from the size of the
individual sources.

    In fact, this latter correction is countered by the correction from the
singles distributions in Eq.~(\ref{doubles1}).  If we assume, solely as a
mathematically simple example, that $d^3n/d^3K \sim e^{-\xi^2 K^2}$, then
\beq
C(q) = 1+e^{-\vec q\,^2(R_0^2+(R_s^2-\xi^2)/2)}
                e^{-{q^0}^2(\tau_s^2+\tau_0^2/2)}.
\label{corralpha}
\eeq
The net deviations of the measured scales $R$ and $\tau$ from the scales of
the source distributions $R_0$ and $\tau_0$ are of order of a few percent at
most.

    More generally one can define a total-momentum dependent pair source
function,
\beq
S_P(X) = \int dx e^{-iPx/2} \langle J^\dagger(X+x/2)J(X-x/2)\rangle,
\eeq
where $P=p+p'$ is the total four-momentum of the pair.  For the simple
example (\ref{JJfact}), we have $S_P(X)=e^{-X^2/2R^2}g(P/2)$; note the
relation to the description (\ref{pack1}) in terms of wavepackets produced by
the source.  In terms of $S$, the HBT correlation function becomes
\cite{pratt84},
\beq
C(q)= 1+ \frac{|\int dX e^{iqX}S_P(X)|^2}
{\int dX S_{P+q/2}(X)\int dX S_{P-q/2}(X)}.
\label{pratt}
\eeq
This equation relates the HBT correlation function to four-dimensional
Fourier transforms of the source function $S_P(x)$.  Compare with the result
(\ref{cf}), which gives the correlation function in terms of the Fourier
transforms of the wavepackets making up the pion distribution.

    The simplest approximation is to take $\xi_c = \tau_c = 0$, or
equivalently, to ignore the $P$ dependence in $S_P(X)$.  Then the
current-current correlation becomes a function of only one variable:
\beq
\langle J^\dagger(x)J(x')\rangle \to S(x)\delta(x-x').
\eeq
This approximation is excellent for stars, where the emission of a photon
is coherent on the order of an atomic scale, while the size of the star is
$\sim 10^{11}$ cm.  Neglecting the correlation length is not as good an
approximation in a heavy-ion collision.

    With the neglect of the finite size of the correlation lengths, we find
from Eq.~(\ref{singles}) that
\beq
\eps_p\frac{d^3n}{d^3p} = \frac{1}{2} \int dx S(x),
\label{singles1}
\eeq
i.e., the singles distribution is flat in momentum.  Furthermore the pair
distribution (assuming a chaotic source) becomes
\beq
\eps_p\eps_p'\frac{d^6n}{d^3pd^3p'}
= \left| \frac{1}{2} \int dx S(x)\right|^2 +
\left| \frac{1}{2} \int dx S(x)e^{i(p-p')x}\right|^2,
%\nonumber \\
\label{epsS}
\eeq
and
\beq
C(q) = 1+ \frac{\left| \int dx S(x)e^{iqx}\right|^2}
           {\left| \int dx S(x)\right|^2}.
\label{corrfcn}
\eeq

\section{Parameterizations of data}

    The most straightforward way to analyze HBT data is to parametrize the
correlation function $C(q)$ as a Gaussian in $q$.  Expanding $C(q)$ in
Eq.~(\ref{pratt}) for small $q$ to second order we find
\beq
C(q)= &2 -
q^\mu q^\nu(\langle X_\mu X_\nu\rangle
- \langle X_\mu\rangle \langle X_\nu\rangle)
\nonumber \\
&+q^\mu q^\nu(\langle x_\mu x_\nu\rangle
- \langle x_\mu\rangle \langle x_\nu\rangle) +
\cdots,
\label{expq}
\eeq
where
\beq
\langle \theta(x,X) \rangle = \frac{\int dx dX \theta(x,X) \langle
J^\dagger(X+x/2)J(X-x/2)\rangle}{\int dx dX \langle
J^\dagger(X+x/2)J(X-x/2)\rangle}.
\eeq
The terms in $x$ in (\ref{expq}), which come from expansion of the
denominator in Eq.~(\ref{pratt}), are of relative order $(\xi_c/R)^2,
(\tau_c/\tau)^2$, and are often neglected in the interpretation of the data,
although, as mentioned, they can modify the extracted sizes by a few percent.
Dropping these latter terms we have
\beq
C(q)= 2 - q^\mu q^\nu(\langle x_\mu x_\nu\rangle
- \langle x_\mu\rangle \langle x_\nu\rangle) + \cdots
\label{expq2}
\eeq
where here and below $\langle \theta(x) \rangle = \int dx S_P(x)\theta(x)/\int
dx'S_P(x')$.  This form suggests a parame\-trization~\cite{yano,podgor},
\beq
C(q)= 1 + \lambda e^{-q^\mu q^\nu(\langle x_\mu x_\nu\rangle
- \langle x_\mu\rangle \langle x_\nu\rangle)},
\eeq
where we also introduce the {\it chaoticity parameter} $\lambda$; for a
completely chaotic source the correlation function rises up to 2 as $q\to 0$,
and thus $\lambda$ = 1, while for a completely coherent source, such as a
laser, $\lambda$ = 0.

    Various increasingly sophisticated versions of this parametrization
have been adopted.  The simplest is to write
\beq
C(q) = 1+\lambda e^{-Q_{\rm inv}^2 R^2},
\eeq
where $Q_{\rm inv}$ is the invariant momentum difference of the two
particles, $Q_{\rm inv}^2 = (\vec p_1 - \vec p_2)^2 - (\eps_{p_1} -
\eps_{p_2})^2$.  Results of such a single-size analysis by NA44 for pairs of
$\pi^+$ and pairs of $\pi^-$ produced in 200 GeV/A collisions of S on Pb at
the SPS \cite{NA44SPb} are shown in Fig.~1, and by E877 for pairs of $\pi^+$,
pairs of $\pi^-$, and pairs of K$^+$ produced in collisions of 10.8 GeV/A Au
on Au at the AGS \cite{e877c} in Fig.~2.  This parametrization corresponds to
the assumption that $\langle x_\mu x_\nu\rangle - \langle x_\mu\rangle \langle
x_\nu\rangle = g_{\mu\nu}R^2$.  Since the sign of the contribution of the
time-time component should be the same as the space-space components a
somewhat better single-size parametrization would be to assume that $\langle
x_\mu x_\nu\rangle - \langle x_\mu\rangle \langle x_\nu\rangle =
\delta_{\mu\nu}R^2$.  Then
\beq
C(q) = 1+\lambda e^{-(\vec q\,^2+{q^0}^2)R^2};
\eeq
cf.  Eq.~(\ref{corralpha}).

    A second level of approximation is to distinguish the space and time
dependence of the evolving system, taking a spherical fireball in space, so
that
\beq
C(q) = 1+\lambda e^{-(\vec q\,^2 R^2 +{q^0}^2\tau^2)}.
\eeq
The time $\tau$ is essentially the duration of the collision:  $\tau^2 =
\langle t^2\rangle - \langle t \rangle^2$, and $R$ the radius of the collision
volume:  $R^2 = \langle \vec r\,^2\rangle - \langle \vec r\, \rangle^2$.

    The next level is to try to take the evolving geometry into account,
including non-sphericity of the source and possible flow effects.  Consider a
pair of particles of total three-momentum $\vec P$ and relative three-momentum
$\vec q$.  Since $q\cdot P = (p-p')\cdot (p+p') = p^2 - p'^2 = 0$, we have
\beq
q^0 = \vec q\cdot\vec P/P^0 = \vec q\,\cdot \vec v\,,
\eeq
where $\vec v$ is the velocity of the center-of-mass of the pair of particles,
$\vec P/P^0$. Then
$q^\mu x_\mu = \vec q\,\cdot (\vec r\, -\vec v\,t)$, and
\beq
q^\mu q^\nu(\langle x_\mu x_\nu\rangle -  \langle x_\mu\rangle \langle
x_\nu\rangle)
= \langle \left( \vec q\,\cdot (\vec r\, -\vec v\,t)\right)^2\rangle
- \langle  \vec q\,\cdot (\vec r\, -\vec v\,t)\rangle^2.
\eeq
Let us erect a three dimensional coordinate system in which the {\it
longitudinal} direction is along the beam axis, the {\it outwards} axis (the
$x$ direction) is along the transverse component of $\vec P$, and the third,
or {\it side}, axis is in the $y$ direction.  In this frame $v_y$ vanishes.
(Note that this coordinate system varies with the pair of particles studied.)
The ensemble of events is symmetric under $y\to -y$, so that cross terms
involving a single $y$ vanish.  However, the cross terms $\langle zt \rangle -
\langle z \rangle\langle t \rangle$ and $\langle xt \rangle - \langle x
\rangle\langle t \rangle$ are generally non-zero, and we find
\beq
\lefteqn{q^\mu q^\nu(\langle x_\mu x_\nu\rangle
  -\langle x_\mu\rangle \langle x_\nu\rangle)}
\nonumber \\
=& q_{\rm out}^2 (\langle (x-v_xt)^2\rangle - \langle x-v_xt\rangle^2)
+ q_{\rm side}^2 (\langle y^2\rangle - \langle y\rangle^2)
\nonumber\\
&+ q_{\rm long}^2 (\langle (z-v_zt)^2\rangle - \langle z-v_zt\rangle^2)
\nonumber\\
&+2q_{\rm out}q_{\rm long}(\langle (x-v_xt)(z-v_zt)\rangle -
\langle x-v_xt\rangle \langle z-v_zt\rangle),
\eeq
a form which suggests a parametrization of the correlation function in
terms of four radii \cite{chapman},
\beq
C(\vec q\,) = 1+\lambda e^{-(q_{\rm out}^2R_{\rm out}^2 +
q_{\rm side}^2R_{\rm side}^2+q_{\rm long}^2R_{\rm long}^2
+2q_{\rm out}q_{\rm long}R_{\rm ol}^2)},
\eeq
where the parameters have the interpretation
\beq
R_{\rm out}^2 &=& \langle (x-v_xt)^2\rangle - \langle x-v_xt\rangle^2,
\nonumber \\
R_{\rm side}^2 &=& \langle y^2\rangle - \langle y\rangle^2,
\nonumber \\
R_{\rm long}^2 &=& \langle (z-v_zt)^2\rangle - \langle z-v_zt\rangle^2,
\nonumber \\
R_{\rm ol}^2 &=&
\langle (x-v_xt)(z-v_zt)\rangle - \langle x-v_xt\rangle \langle z-v_zt\rangle.
\label{3d}
\eeq
Note that $R_{\rm ol}^2$, although written as a square, need not be
positive.  It is often convenient to analyze data, pair-by-pair, in the
``longitudinal center-of-mass" frame, in which $P_z = 0$; then
$q_{\rm out}=q^0/v$, and $R_{\rm long}^2$ reduces to $\langle z^2\rangle -
\langle z\rangle^2$.

    The {\it three-dimensional} parametrization, Eq.~(\ref{3d}), is commonly
used in interpreting present HBT mesurements.  Typical three-dimensional
analyses of correlations of pion pairs are shown in Figs.~9 and 10, in Fig.~9
correlations of of $\pi^+ \pi^+$ from 200 GeV/A S+Pb collisions studied by
NA44~\cite{NA44-3dSPb}, and in Fig.~10 correlations of $\pi^+ \pi^+$ and
$\pi^- \pi^-$ from 10.8 GeV/A Au+Au collisions studied by E877~\cite{e877b}.

    Considerable information on the development of the collision volume, e.g.,
flow \cite{voloshin}, can be extracted from the three-dimensional analyses.
The experimental dependence of the two-particle correlations on the momenta of
the particles indeed indicates that the systems are expanding.  For detailed
discussions of the physics extracted from recent experiments, see, e.g., Refs.
\cite{heinz96a} and \cite{sakaguchi}.

\begin{figure}
\begin{center}
\epsfig{file=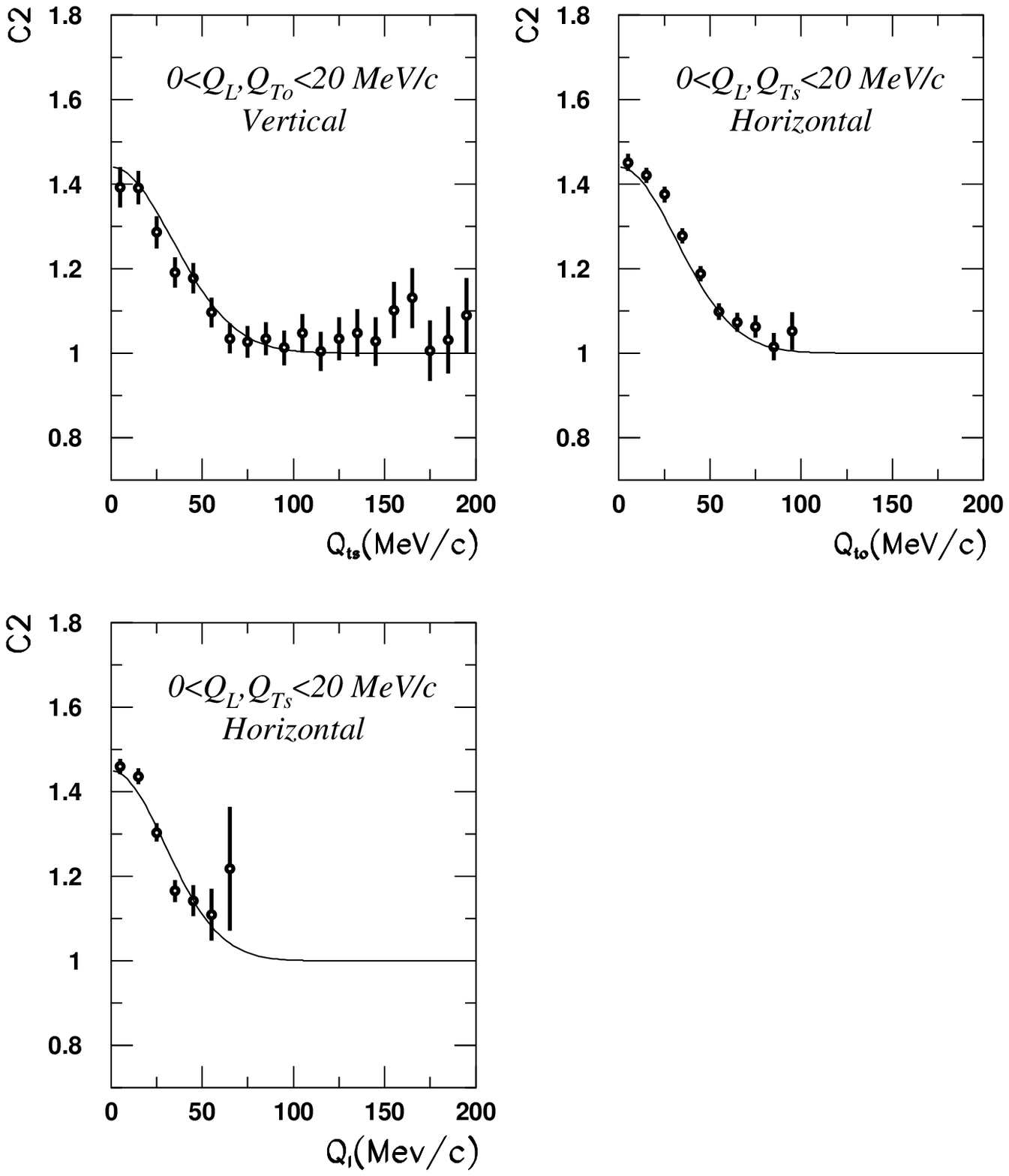, height=11cm}
\end{center}
Fig.~9. Three-dimensional fittings of $\pi^+ \pi^+$ correlation
functions, from NA44~\cite{NA44-3dSPb}, the upper left panel as a function of
$q_{\rm side}$, the upper right as a function of $q_{\rm out}$, and the lower
panel as a function of $q_{\rm long}$.
\end{figure}

\begin{figure}
\begin{center}
\epsfig{file=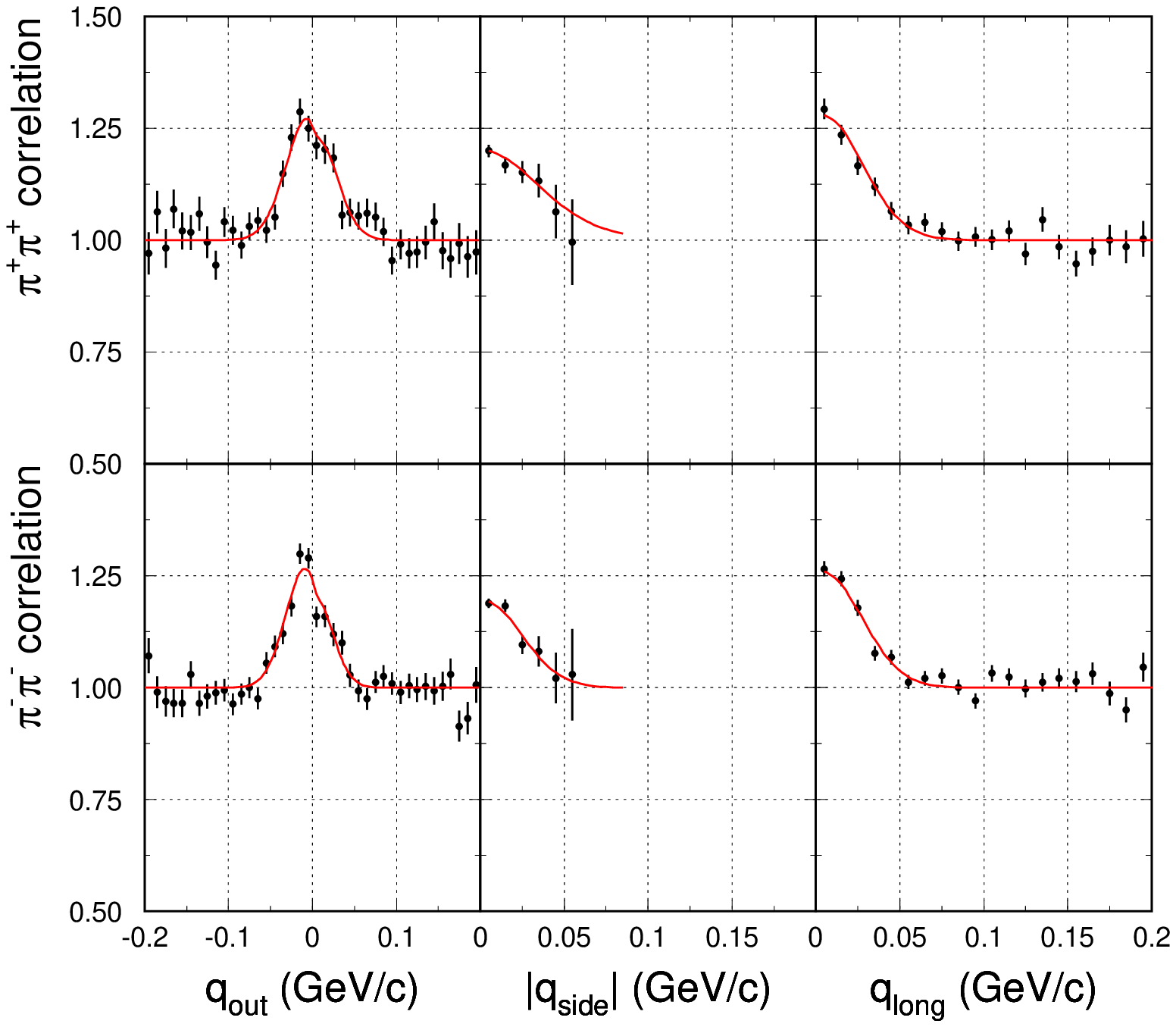, height=10cm}
\end{center}
Fig.~10.  Three-dimensional fittings of $\pi^+ \pi^+$ and $\pi^- \pi^-$
correlation functions, from E877~\cite{e877b}.
\end{figure}

\begin{figure}
\begin{center}
\epsfig{file=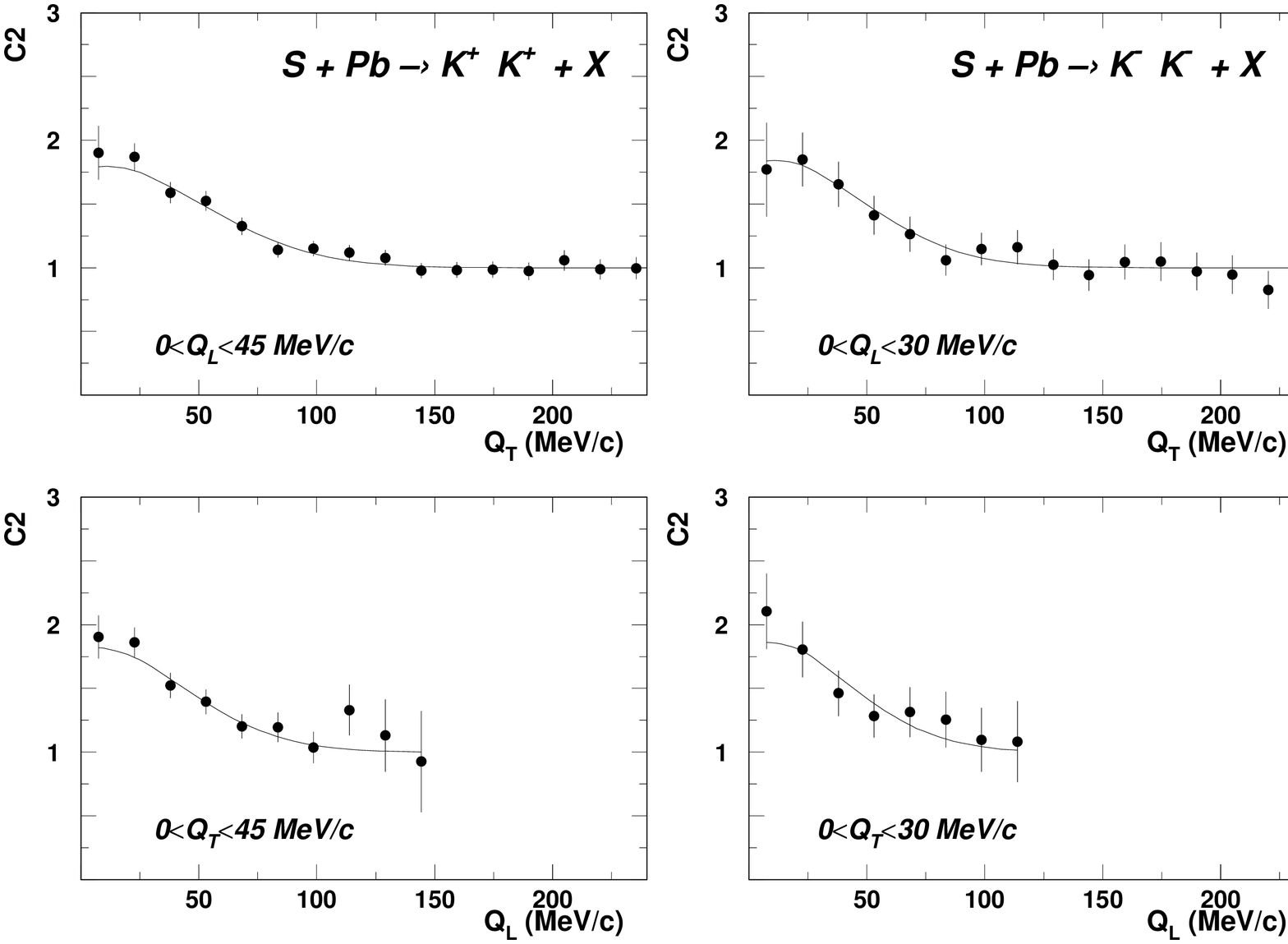, height=9cm}
\end{center}
Fig.~11.  Two-dimensional fittings of K$^+$K$^+$ (left panels) and K$^-$
K$^-$ (right panels) correlation functions, as functions of $q_{\rm transv}$
and $q_{\rm long}$.  From Ref. \cite{jacak1}.
\end{figure}

    Figure~11 shows NA44 data on K$^+$K$^+$ and K$^-$K$^-$ correlations,
projected as functions of $q_{\rm long}$ and the component $q_{\rm transv}$ of
$\vec q\,$ perpendicular to the beam axis~\cite{jacak1}.  Note that the
chaoticity parameter $\lambda$ is somewhat larger for kaons than pions, a
point which examine in the following section.

\section{Sources of chaoticity $\lambda < 1$}

    The chaoticity parameter, $\lambda$, is generally found experimentally to
be less than one, a reflection of intrinsic physical effects as well as
experimental difficulties.  The most fundamental effect would be that the
source exhibits a level of coherence, the situation in a laser, or a form of
pion or other boson condensate.  HBT measurements of pions produced from a
disordered chiral condensate in an ultrarelativistic heavy-ion collision would
also show a reduced $\lambda$~\cite{dcc}.  However, as we noted above,
rescattering by other particles in the collision volume tends to destroy phase
correlations from the production process.  Another example is the MIT atom
laser \cite{atomlaser} where magnetically trapped and evaporatively cooled
sodium atoms are extracted in coherent states from a Bose-Einstein condensed
system; since the extracted atoms do not exhibit an HBT effect, $\lambda$
would be zero.  (See the discussion of HBT in atomic beams below.)

    Even if the source is completely chaotic, measurements do not necessarily
give $\lambda$ =1.  The first reason is the simple but important problem of
contamination of the sample from misidentification of particles, e.g., an
e$^-$ or K$^-$ as a $\pi^-$, so that one includes pairs of non-identical
particles in the data set.

    A second stems from unravelling the effects of Coulomb final state
interactions between a pair of identical charged particles.  The point is that
the methods of removing effects of Coulomb interactions, which we discuss in
some detail in the following section, become more uncertain the smaller the
relative momentum difference, leading to uncertainty in $\lambda$.

    The next physical effect reducing $\lambda$ is the production of pions
from long-lived resonances.  Such pions appear to come from sources of large
radii, which would give an HBT enhancement only at very small $q$.  Indeed,
perhaps half of the pions produced in an ultrarelativistic heavy-ion collision
come from resonances, rather than being produced directly.  Pions from
short-lived resonances, e.g., from $\rho \to \pi\pi$ or from $\Delta \to N
\pi$, are produced well within the collision volume and are not an issue.  On
the other hand, the long-lived resonance $\eta$ has a lifetime of order 1.2
{\AA}/c, and the 3$\pi$ into which it decays would appear to be produced at a
relatively enormous distance of order {\AA} from the collision volume.  The
$\omega$ goes some 24 fm, the $\eta'$ some 800 fm, etc.  The result is that
the collision volume is surrounded by a halo of pions from resonances.

    A small chaotic source would lead to a broad HBT correlation function,
while a very large source would lead to a correlation function with a sharp
rise only at small relative momenta.  Now if one has both a small and a large
source, e.g., a partially transparent cloud in front of the sun, one sees a
combination of both, as illustrated in Fig. 7c, where the width of the bump
closest to the origin is inversely proportional to the size of the large
source (the cloud), and the width of the broader bump reflects the size of the
small source (the sun).  Figure 12 illustrates how pions from different
resonances contribute to the correlation function, here as a function of the
{\it out} component of the momentum difference, in a central CERN S-Pb
collision~\cite{HHres}.  The estimate is based on an RQMD simulation of the
collision.  Note the rise at very small relative momenta from long-lived
resonances.\footnote{The calculation assumes that the pions from resonances
are described by the same factorized form of the two-pion correlation function
as the pions emerging directly from the collision volume.  However, the pions
from longer-lived resonances do not undergo any rescattering and thus reflect
the statistics of the source resonances, which can in principle decrease the
contribution to the HBT effect at small relative momenta.} However, unless one
is capable of resolving the little peak at small $q$ one would deduce that the
data goes to a value of $\lambda$ less than one.  The effect on the observed
$\lambda$ is a reduction of $\sim$ 30\% for pions, and $\sim$ 10\% for kaons,
since a smaller fraction of kaons are produced by long-lived resonances.
Since the production of resonances falls off with increasing transverse
momentum, one finds in fact that the contribution of pions from resonances to
the HBT signal decreases as $p_\perp$ of the pions studied increases.

    In the simple model of HBT described above we assumed well defined wave
functions propagating through vacuum.  But in reality the particles propagate
through of order 1 mm of target and then, e.g., in NA44, through 15 meters of
air.  Let us consider the effects of secondary scattering in the target and
the intervening air.  From a quantum mechanical point of view scattering from
the target and air atoms changes an initial pure state wave function of a
particle into a mixed quantum state:  when a pure wave function hits the atoms
in the target or air, it generates, \`a la Huygens, a beam of secondary wave
functions; because the atoms are disturbed, the secondary waves become
incoherent with the initial wave.  The secondary interactions produce many
small angle scatterings of the particles, which have the observational effect
of distorting the correlation function \cite{air}.

\begin{figure}
\begin{center}
\epsfig{file=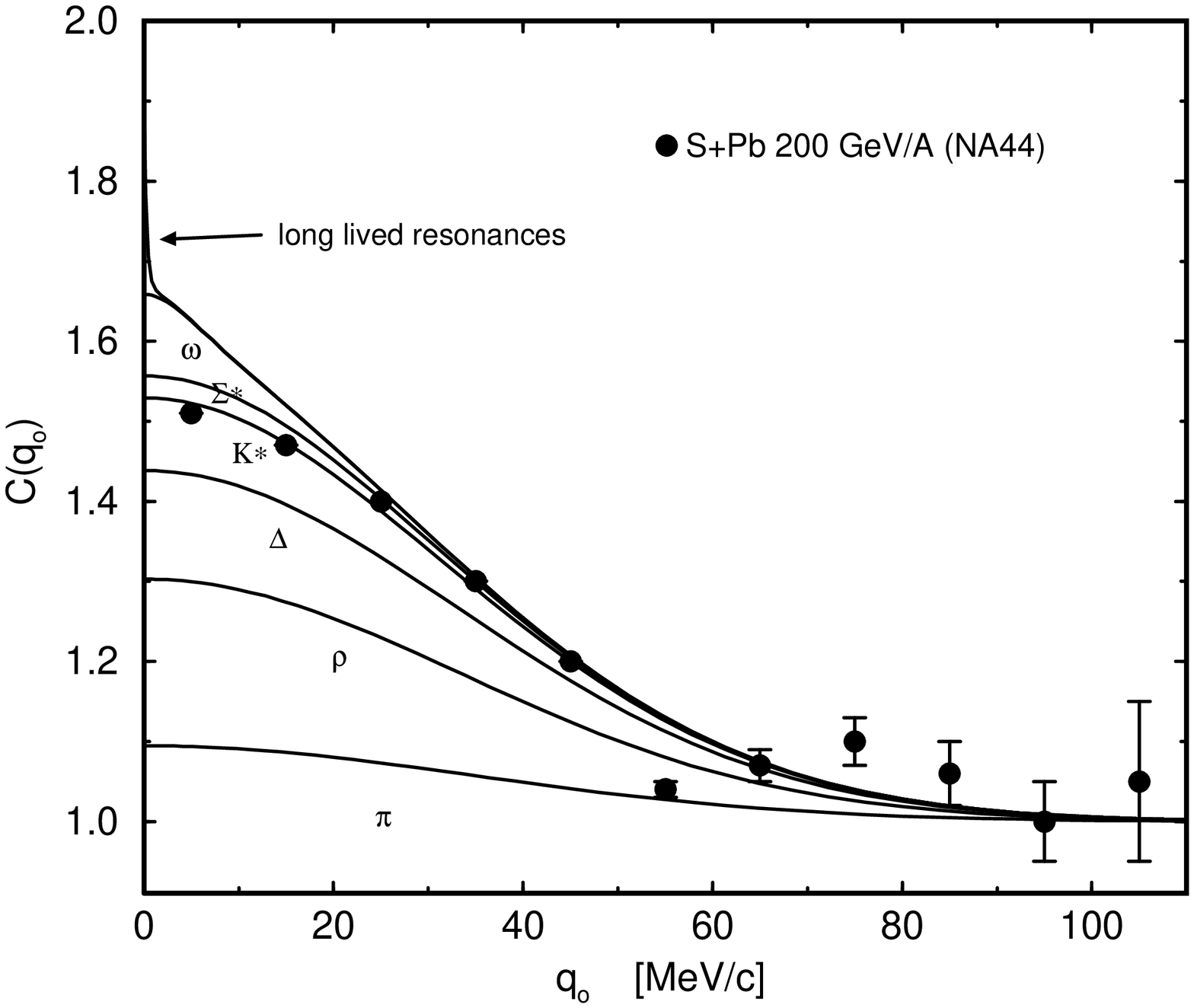, height=8cm, angle=-90}
\end{center}
Fig.~12.  Estimated contributions of pions from resonances in HBT compared
with NA44 data~\cite{NA44SPb}. From~\cite{HHres}.
\end{figure}

    To see the effects of small angle scattering, it is adequate to describe
the interaction of a high energy particle with an atom of charge $Z$ by a
screened Coulomb potential, $V(r) = Ze^2 e^{-r/a}/r$, where $a =
0.8853a_0/Z^{1/3}$ is the Fermi-Thomas radius of the atom and $a_0$ is the
Bohr radius, and neglect elastic scatterings due to strong interaction with
the nucleus.  The differential cross section of an incident particle of
momentum $p$ is given in the Born approximation by
\beq
\frac{d\sigma}{d\Omega} = \frac{4p^2Z(Z+1)\alpha^2}{(q^2+1/a^2)^2},
\label{screen}
\eeq
where $\vec q\,$ is the momentum transfer in the scattering, and
the substitution $Z\to Z+1$ takes into account scattering by atomic
electrons~\cite{bethe}.  While the total cross section is $\sigma = 4\pi
Z(Z+1) \alpha^2 a^2$, the effects of multiple scattering are more accurately
controlled by the transport cross section, $\sigma_t \equiv \int d\Omega
(1-\cos \theta)d\sigma/d\Omega $, given here by
\beq
\sigma_t = \frac{4\pi Z(Z+1) \alpha^2}{p^2} (\ln(2pa) - 1/2).
\eeq
In a single scattering, $\langle \cos \theta \rangle = 1- \sigma_t/\sigma$, so
that at high energies,  $\langle \theta^2 \rangle \simeq 2\sigma_t/\sigma$.

    Consider a particle going a distance $L$ through a medium of atomic
density $n_a$.  From the multiple scattering equation one readily finds that
the particle emerges with $\langle \cos \theta \rangle = e^{-n_aL\sigma_t}$,
and more generally, with $\langle {\rm P}_\ell(\cos \theta) \rangle =
e^{-n_aL\sigma_\ell}$, where $\sigma_\ell = \int d\Omega (1-{\rm P}_\ell(\cos
\theta)) d\sigma/d\Omega $ \cite{goudsmit}.  The underlying angular
distribution is more complicated, but for our present estimates we may assume
that the spread in angles is Gaussian:  $f(\theta,L)\theta d\theta \sim
e^{-\theta^2/\langle \theta^2\rangle} \theta d\theta$.  The mean square
scattering angle $\langle \theta^2\rangle$ is given by the non-linear
equation~\cite{bethe},
\beq
\langle \theta^2\rangle - \frac{4\pi Z(Z+1)\alpha^2}{p^2} n_aL \ln
\langle \theta^2\rangle = 2n_aL\sigma_t^{\rm eff} \equiv
\langle \theta^2\rangle_0,
\label{Moliere}
\eeq
where
\beq
\sigma_t^{\rm eff} =  \frac{4\pi Z(Z+1) \alpha^2}{p^2} \ln (pa/\nu^{1/2}),
\eeq
and the factor $\nu \approx 1.32(1+3.34Z^2\alpha^2)$ includes corrections
to the cross section beyond the Born approximation.  The second term on the
left in Eq.~(\ref{Moliere}) approximately takes into account effects of
large angle scatterings, and reduces $\langle \theta^2\rangle$ from $\langle
\theta^2\rangle_0$.

    Let us consider as illustration the effect on a 4 GeV pion scattering
through 1 mm of Pb.  Then the mean scattering angle is $\sim 2 \times
10^{-3}$, which produces a mean transverse spread in momentum, $\Delta
p_\perp$, of order 8 MeV, corresponding to a transverse deflection, $\Delta
r_\perp$, of order 3 cm.  Similarly, a 4 GeV pion traversing 15 m of air
($Z=7$) undergoes a mean angular deflection $\sim 0.9 \times 10^{-3}$, with
$\Delta p_\perp \sim 3.7$ MeV and $\Delta r_\perp \sim 1.4$ cm.  The pion
makes some $10^3$ scatterings per cm; air is remarkably opaque to pions.

    The effect of these small angle deviations is to spread out the singles
distributions and the correlation function.  Such secondary scattering effects
are generally accounted for in the estimated experimental momentum
resolution.  An initial distribution of single particle transverse momenta,
$n^0_{\vec p_\perp}$, will be spread into a final distribution,
\beq
n_{\vec p_\perp} \sim \int e^{-(\vec p_\perp - \vec p_\perp^\prime)^2/\Delta
   p_\perp)^2} n^0_{\vec p_\perp^{\,\prime}} d^2 \vec p\,_\perp^\prime.
\eeq
Similarly, two particles starting out with a given relative momentum and
undergoing random walks do not end up with the same final relative momentum.
For example, two particles detected with zero relative momentum may have
actually started out at a larger momenta and have been bent in by the air or
target.  An initial HBT correlation function $C_0(q) = 1 + e^{-\vec
q\,^2R_0^2}$ will be spread into an observed correlation function,
\beq
C_{\rm obs}(q) = 1 + \lambda_{\rm eff} e^{-\vec q\,^2R_{\rm eff}^2},
\eeq
where the measured radius $R_{\rm eff}$ is decreased from the original radius,
$R$, by a factor
\beq
\frac{R_{\rm eff}}{R} = \frac{1}{[1+2(R\Delta p_\perp)^2]^{1/2}},
\eeq
and the chaoticity parameter is reduced from unity to
\beq
\lambda_{\rm eff} = \left(\frac{R_{\rm eff}}{R}\right)^2.
\eeq
For example, with an initial nominal radius of $R$ = 7 fm, scattering in
air reduces the measured $R$ by 2\% and the measured $\lambda$ by
3\%.  Including the effect of a 1 mm Pb target, one finds a 7\%
reduction in the measured $R$ and a total reduction of $\lambda$ due
to secondary scattering of 16\%.  The effects on HBT of secondary
scattering in a thick target can be substantial; e.g., for 1 cm of Pb,
$\lambda$ falls below 0.4.

    The astute reader may at this point have noticed a contradiction between
the picture of secondary scattering in a cloud obscuring a smaller source, the
sun say, leading to a narrower correlation function than the one that would be
produced by the sun (cf.  Fig. 7b), and the present picture of scattering in
air, which broadens the correlation function and reduces it at the origin.
I leave the resolution of this problem as an instructive exercise.

\begin{figure}
\begin{center}
\epsfig{file=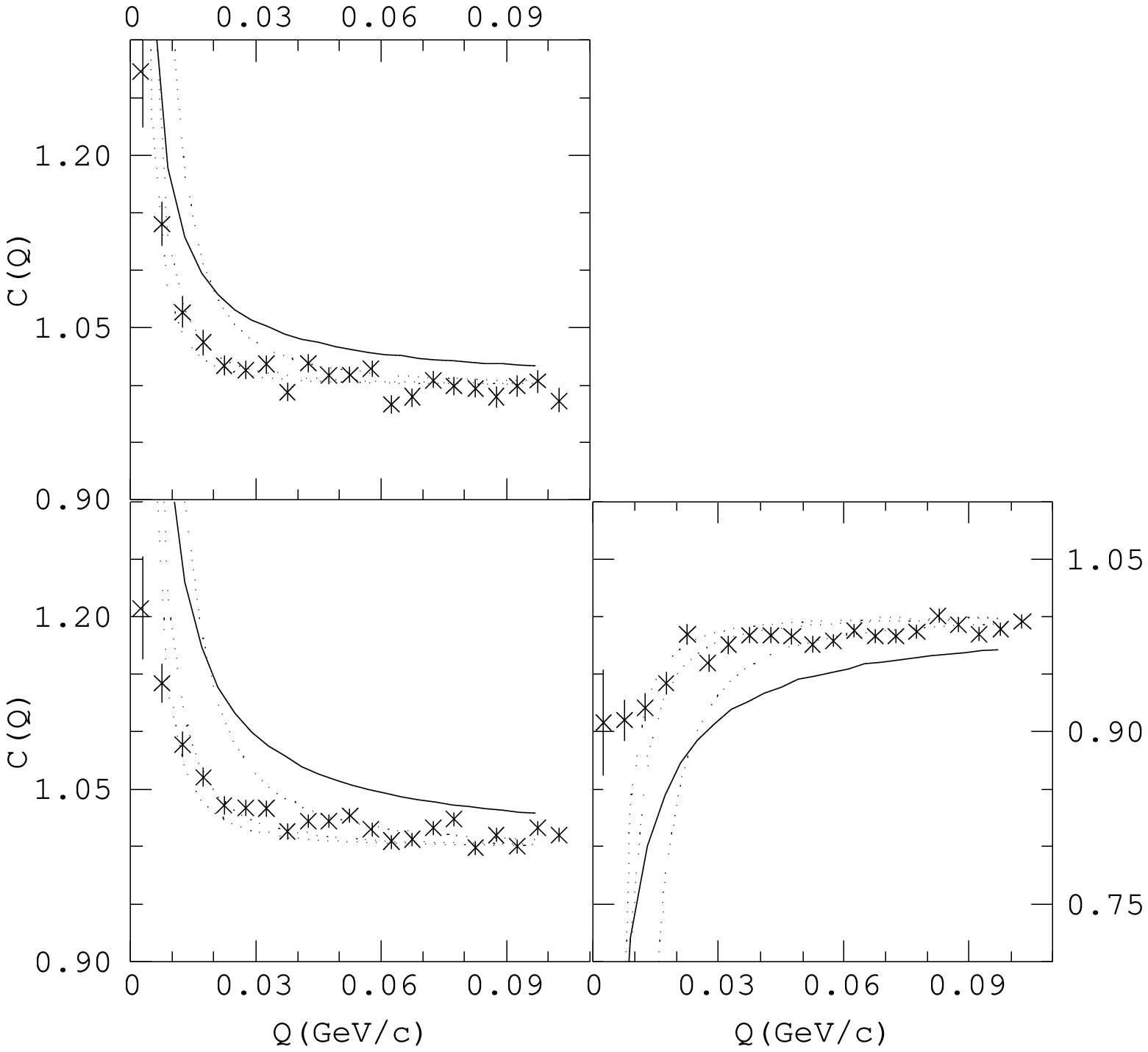, height=10cm}
\end{center}
Fig.~13.  E877 data \cite{e877a} for (a) $\pi^+\pi^-$, (b) $\pi^-p$, and
(c) $\pi^+p$, as well as comparison (dotted lines) with the toy model,
Eq.~(\ref{toydist}) for $r_0$ in the range 3-15 fm, assuming a bare
correlation function $C_0 = 1$, and the Gamow correction (solid line).
\end{figure}

\section{Final state Coulomb interactions}

    Up to now we have discussed Hanbury Brown--Twiss interferometry assuming
that the particles travel completely independently once they leave the
collision region.  In fact one measures correlations primarily among charged
mesons; the Coulomb interactions between any given pair of particles, as well
as those of the individuals in the pair with the other charged particles in
the system, produce important effects on the measured correlation functions.
Even though the detectors are many meters from the collision, those that
produce the enhanced signal are typically within a meter of each other.  The
pions whose correlation one measures travel essentially along the same
direction and they continue to have a strong Coulomb interaction the entire
way.  Disintangling these final state Coulomb interactions is a very difficult
question.\footnote{Most of the material described in this section was
developed by the author and P. Braun-Munzinger, with J. Popp's helpful
assistance, and reported in Ref.  \cite{hbtpbm}.}

    Bare data, uncorrected for Coulomb interactions, brings out the situation
very clearly.  Figure 13 shows the uncorrected E877 correlation function
measurements for $\pi^+\pi^-$, $\pi^-p$, and $\pi^+p$ pairs produced in Au+Au
collisions at the AGS at 10.8 GeV per nucleon \cite{e877a}, while Fig. 14
shows bare data for the $\pi^+\pi^+$ and $\pi^-\pi^-$ correlation functions.
The correlation function for distinguishable particle pairs does not have the
expected value of unity, while the correlation function for identical
particles does not rise up anywhere as high as in the Coulomb-corrected data,
e.g., Figs. 1, 2, and 9--11.  Rather, the data for both identical particles
and oppositely charged non-identical particles are very similar.

    The traditional method of correcting for final state Coulomb interactions
is to employ the Gamow correction.  For example, in the beta decay of a
neutron into a proton, electron, and antineutrino, the proton and electron are
produced in a relative Coulomb state.  Because the electron and proton are
attracted to each other, the amplitude for them to be at the origin is
enhanced.  The decay amplitude is a bare matrix element times the relative
electron-proton Coulomb wave function, $\psi_C(0)$, at the origin.
Non-relativistically, the Coulomb wave function for the relative motion of
a pair of particles of charges $z e$ and $z^\prime e$ with relative momentum
$\vec Q = (\vec p\, - \vec p\,^\prime)/2 = {\vec q}/2$ at infinity, and
relative velocity $v_{\rm rel} = Q/m_{\rm red}$, is $\psi_C(\vec r\,) =
\psi_C(0) _1F_1(-i\eta;1;i(Qr-{\vec Q}\cdot {\vec r}))$, where the
dimensionless parameter $\eta(Q)$ is given by
\beq
  \eta = \frac{zz^\prime \alpha }{v_{rel}/c},
\label{eta}
\eeq
and the reduced mass $m_{\rm red}$ equals $m/2$ for two particles of mass
$m$.  Also,
\beq
   \psi_C(0) = \left(\frac{2\pi\eta}{e^{2\pi\eta}-1}\right)^{1/2},
\label{psi0}
\eeq
The actual rate is that which one would measure in the absence of any
Coulomb effects times the Coulomb correction, $|\psi_C(0)|^2$.  For particles
of opposite charge the probability is enhanced, by a factor tending to
$2\pi|\eta|$ at small $Q$.  On the other hand imagine a decay of a
$\Delta^{++}$ into a proton, positron and neutrino.  The proton and positron
repel each other, and thus have a reduced amplitude to be at the origin.  The
Coulomb correction, $|\psi_C(0)|^2$, is less than unity in this case, and the
net rate would be suppressed from its value in the absence of Coulomb
interactions, by a factor tending to $2\pi\eta e^{-2\pi\eta}$ at small $Q$.

    In making a Gamow correction in heavy-ion collisions, one assumes that the
pair of identical particles is produced in a relative Coulomb state at zero
separation, and thus the amplitude for doing so is reduced from the bare
amplitude by the factor $\psi_C(0)$.  The {\it bare} correlation function,
$C_0(q)$, where $q=2Q$, is thus extracted from the measured correlation
function, $C(q)$, by dividing out the assumed factor $|\psi_C(0)|^2$ in the
production rate:
\beq
C_0(q) = \frac{C(q)}{|\psi_C(0)|^2} =
      C(q)\frac{(e^{2\pi\eta(Q)}-1)}{2\pi\eta(Q)}.
\label{gamow}
\eeq
Note that as the relative momentum goes to zero, $\eta \to +\infty$ and
$\psi_C(0) \to 0$ for identical particles, and Eq.  (\ref{gamow}) yields an
infinite correction.

    The question is why one should assume that the Coulomb wave function at
the origin should control the Coulomb corrections?  For particles of the same
charge, $\psi_C(r)$ falls to zero exponentially as the particles approach each
other inside the (zero angular momentum) {\it classical turning point}, $r_t$,
defined by $q^2/2m_{\rm red} = e^2/r_t$.  Outside $r_t$ it oscillates, and
describes essentially classical physics.  In order for the physics at the
origin to be relevant, it is necessary that the source be highly localized
compared to the distance to the turning point.  However, for pions in a
heavy-ion collision, $r_t\simeq (200$ fm)/$Q^2$, where $Q$ is measured in
MeV/c; for $Q\sim$ 10 MeV/c, a typical minimum value, $r_t$ is only 2 fm, and
smaller for larger $Q$.  Since $r_t$ is much smaller than the characteristic
heavy-ion radius, most of the pairs of particles observed in a heavy-ion
collision are in fact made at relative separations well outside their
classical turning points.

\begin{figure}
\begin{center}
\epsfig{file=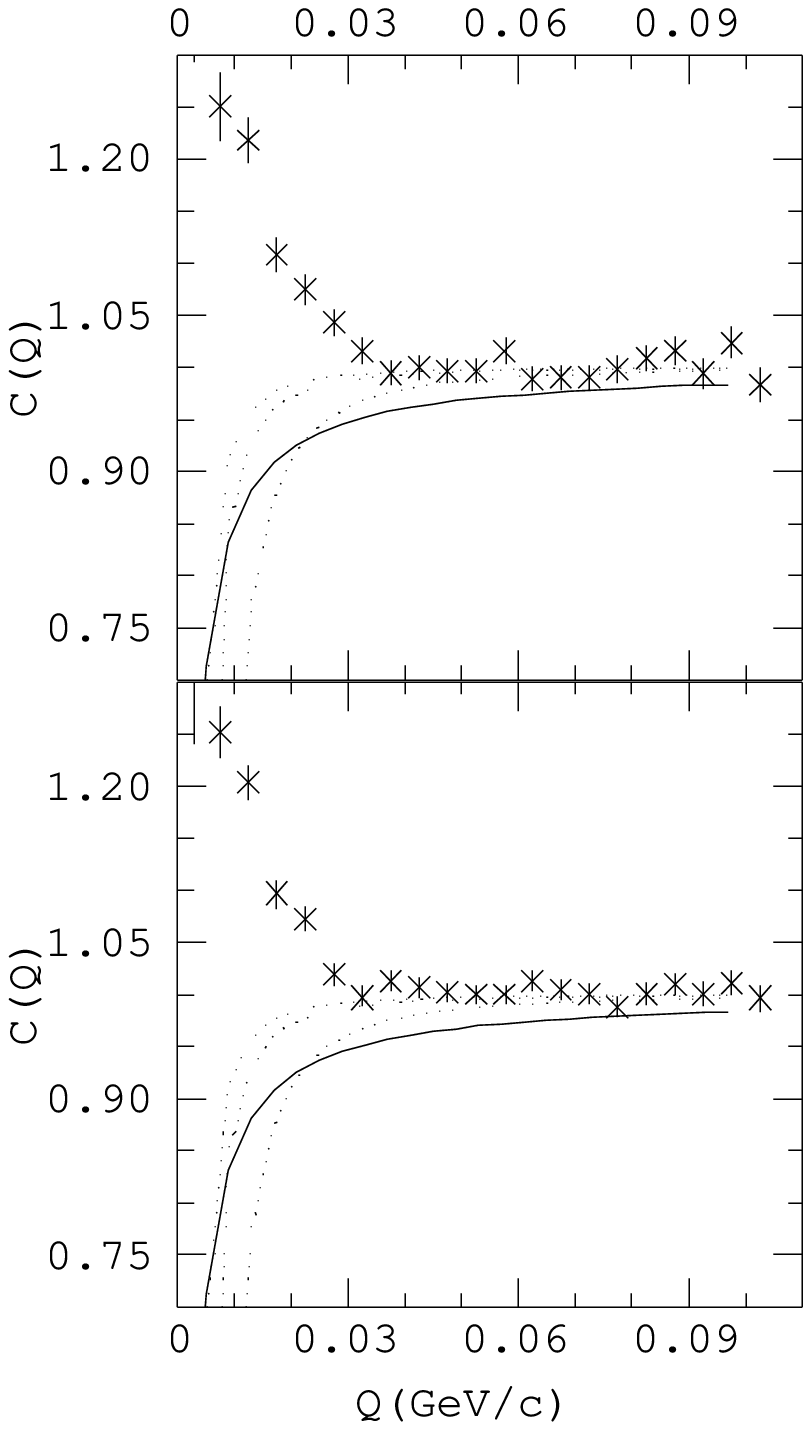, height=10cm}
\end{center}
Fig.~14.  E877 data \cite{e877a} for $\pi^+\pi^+$ and (b) $\pi^-\pi^-$,
together with the Coulomb correction, Eq.  (\ref{toydist}), for the same range
of $r_0$ as in Fig. 13.
\end{figure}

    There are three relevant length scales in the Coulomb problem, the
classical turning point, $r_t$, the wavelength of the relative motion, and the
two-particle Bohr radius, $a_0 = 1/m_{\rm red}e^2$ (which is 387 fm for
$\pi\pi$ and 222 fm for $\pi$p).  For typical $Q$, these length scales are
cleanly separated:
\beq
 r_t:1/Q:a_0 = 2: a_0Q: (a_0Q)^2.
\label{scales}
\eeq
For $\pi\pi$ (or $\pi$p), $a_0Q = 1/|\eta| = 1.96$ (or 1.13) $Q/({\rm
Mev/c})\gg 1$.  The classical turning point is thus the relevant scale for
Coulomb effects.  These arguments suggest that the Coulomb corrections are
dominated by classical physics.

    The major effect of the Coulomb interaction between the particles in the
pair, at distances large compared with $r_t$, is to accelerate them relative
to each other.  Particles of the same charge are accelerated to larger
relative momenta, thus depressing the observed distribution at small $Q$,
while particles of opposite charge are reduced in relative momentum in the
final state, which builds up the distribution at small $Q$.  Although these
effects are qualitatively similar those produced by the Gamow correction, they
are quantitatively rather different.

    In the presence of many produced particles, the relative motion of the
particles in the pair is strongly affected by their interactions with the
plasma of other particles.  The mutual Coulomb interaction of the pair becomes
dominant only when the pair has sufficiently separated from the other
particles in the system that there is small probability of finding other
particles between the particles in the pair.

    One can write down a simple toy model to take these effects into account,
by simply neglecting the Coulomb interaction between the pair for separations
less than an initial radius $r_0$, and for separations greater than $r_0$
including only the relative Coulomb interaction.  Since the relative motion is
in the classical region, conservation of energy of the pair implies that the
final observed relative momentum $Q$ is related to the initial momentum of the
pair $Q_0$ at $r_0$ by
\beq \frac{Q^2}{2m_{\rm red}} =
\frac{Q_0^2}{2m_{\rm red}} \pm \frac{e^2}{r_0},
\label{toy}
\eeq
where the upper sign is for particles of like charge, and the lower for
particles of opposite charge.  For example, for pions with $r_0$ = 10 fm, $Q^2
= Q_0^2 \pm 20(\rm{MeV/c})^2$.  The physics can be treated
non-relativistically because in the rest frame of the pair, one is interested
in relative momenta small compared with $mc$.

    Since the Coulomb interaction conserves particles and the total momentum
of the pair, the final distribution $d^6n/d^3pd^3p'$
of relative momenta $Q$ is thus given in terms of the initial distribution of
pairs, $d^6n^0/d^3p_0d^3p_0'$, by
\beq
\frac{d^6n}{d^3pd^3p'}d^3Q= \frac{d^6n^0}{d^3p_0d^3p_0'}d^3Q_0
\label{transf}
\eeq
The Jacobian, with changes in relative angles ignored, is, from Eq.
(\ref{toy}), $d^3Q_0/d^3Q = Q_0/Q$.  Neglecting to good accuracy the effects
of Coulomb interactions on the singles distributions we have
\beq
  C(\vec q\,) = \frac{q_0}{q}C_0(\vec q_0\,)
  = \left(1 \mp \frac{2m_{\rm red}e^2}{r_0 Q^2}\right)^{1/2} C_0(\vec q_0\,).
\label{toydist}
\eeq

    Figure 13 compares the predictions of the toy model, Eq.~(\ref{toydist}),
with the E877 data for $\pi^+\pi^-$, $\pi^-p$, and $\pi^+p$ systems in Au+Au
collisions at the AGS \cite{e877a}, assuming that the bare correlation
function $C_0$ equals unity.  The dashed lines are the results of the toy
model for $r_0$ = 3 fm (rightmost curve), 9 fm, and 15 fm (leftmost curve),
along with the standard Gamow correction (solid line).  Except at very small
relative momenta $Q\,\la\,10$ MeV/c, where effects due to the finite momentum
resolution of the experiment become visible in the data, the model gives a
good account of the data for $r_0$ in the range of 9 - 15 fm.  By contrast,
the Gamow factor considerably overpredicts the data for all $Q$ shown here.
Similar bare data from NA49 at CERN, for $\pi^+\pi^-$ produced in 160 GeV per
nucleon Pb on Pb collisions is also well fit by the toy model with $r_0\sim
10-20$ fm, while again the Gamow correction is too large, as in Fig. 13
\cite{strobele}.

    Note that the raw correlation data for non-identical particles contains
information about the mean separation of pairs when screening effects become
negligible, summarized in the toy model by the parameter $r_0$, which is
possibly $Q$ dependent.

    With the initial radius $r_0$ extracted from the unlike-sign data, one can
then construct the Coulomb correction for like-sign particles.  The Coulomb
correction factor deduced from Eq.~(\ref{toydist}) is shown for $\pi^+\pi^+$
in Fig. 14a and $\pi^-\pi^-$ in Fig. 14b, for the same range of $r_0$ (as in
Fig. 13, the rightmost curve corresponds to $r_0$ = 3 fm).  Again we see that
use of the Gamow factor implies a correction which differs significantly from
that of the toy model.

\begin{figure}
\vspace{-2cm}
\begin{center}
\epsfig{file=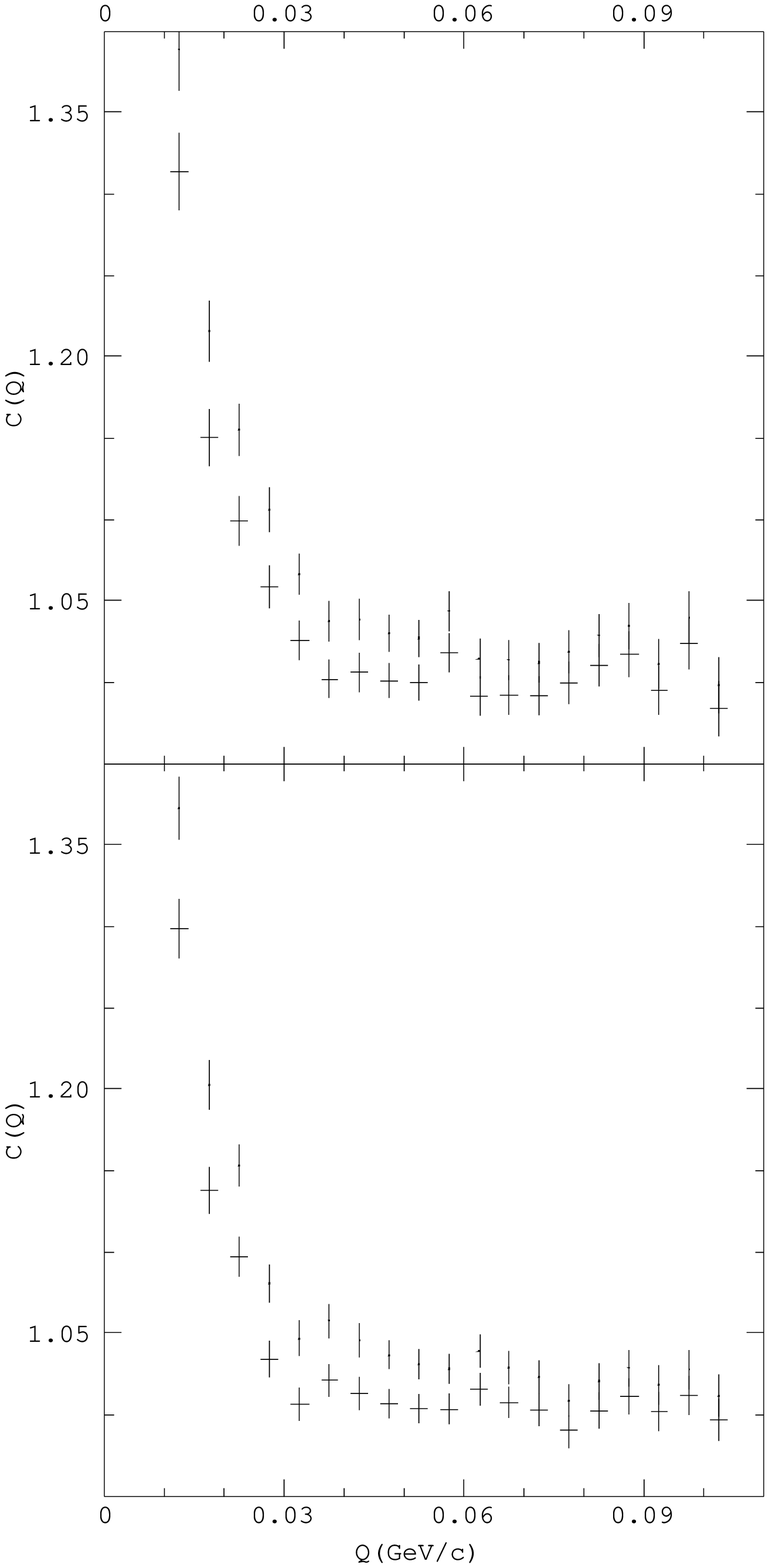, height=16cm}
\end{center}
Fig.~15.
    Toy model calculation of $C(Q)$ for like-sign pions (crosses), compared
with the correlation function derived by making the standard Gamow correction
(vertical bars); (a) $\pi^+\pi^+$ and (b) $\pi^-\pi^-$.
\end{figure}

    Dividing the raw E877 data by the toy model correction factor, with $r_0$
= 15 fm, we obtain the correlation function for like-sign pions (crosses)
shown in Fig. 15a for $\pi^+\pi^+$ and Fig. 15b for $\pi^-\pi^-$, which also
shows the correlation function (vertical bars) derived by making the standard
Gamow correction.  Using the Gamow factor instead of the proper Coulomb
correction leads to a correlation function which is $\sim$ 30\% wider,
implying a correspondingly reduced radius parameter.  Furthermore, the shape
of the ``Gamow-corrected" correlation functions have considerable non-Gaussian
tails in the range $30 < Q < 80 $ MeV/c.  These tails do not exist in the raw
correlation function and obscure the interpretation of the data.

    The length $r_0$ gives one a measure of the scale at which Coulomb
interactions between the particles in the pair dominate their relative motion.
To calculate this decoupling length from microscopic models requires a
non-trivial description of many-particle screening in the high frequency
reqime.  The Coulomb corrections will furthermore change character at RHIC
energies, where the meson density in given events will be sufficiently large
that the Coulomb corrections remain those of a many-particle system out to
much larger distances.

    In order to make a bridge between the toy model and the Gamow correction,
as a first step in constructing a more accurate accounting of Coulomb
corrections, it is instructive to review how the classical description emerges
from the full quantum-mechanical treatment of the Coulomb problem.  In the
absence of Coulomb interactions (denoted by 0 here) the number of pairs of
relative momentum $\vec Q\,$ is given by Eq.~(\ref{d6njjjj}):
\beq
\eps_p\eps_p'\left(\frac{d^6n}{d^3pd^3p'}\right)_0&
\nonumber \\
=& \frac{1}{4} \int dx dx'dx'' dx'''
e^{iP(x+ x''-x' -x''')/2}e^{iQ(x- x''-x' +x''')}
%\nonumber \\ &\times
\langle J^\dagger (x) J^\dagger(x'')J (x''') J(x')\rangle.
\label{d6njjjj0}
%\nonumber \\
\eeq
To take into account the Coulomb interaction only between the pair of
produced particles, we simply replace the relative free-particle wave
functions, $e^{iQ(x-x'')}$ and $e^{iQ(x'-x''')}$, by the Coulomb wave
functions, $\psi_C(x- x'')$ and $\psi_C(x'- x''')$ for the relative motion
for a pair of relative momentum $Q$ at infinity, so that
\beq
\eps_p\eps_p'\frac{d^6n}{d^3pd^3p'}
\nonumber \\
= &\frac{1}{4} \int dx dx'dx'' dx'''
e^{iP(x+ x''-x' -x''')/2} \psi_C(x- x'')\psi_C^*(x'-x''')
%\nonumber \\ &\times
 \langle J^\dagger (x) J^\dagger(x'')J (x''') J(x')\rangle.
\label{d6njjjj1}
%\nonumber \\
\eeq

    Pairs of low relative momentum have relatively low angular momentum, e.g.,
a pair produced at 10 fm separation with relative momentum 20 MeV/c can have
at most one unit of relative angular momentum.  Thus only the low partial wave
components of the Coulomb wave function enter Eq.  (\ref{d6njjjj1}) with
appreciable probability.  Let us consider just s-waves in the WKB
approximation, which is quite good for the s-wave outside the collision
volume.\footnote{The condition for validity of the approximation is $|\partial
p(r)/\partial r| \ll p(r)^2$, which for $r\ll a_0$, the region of interest,
becomes the restriction, $r \ga 3/q^{3/2}a_0^{1/2}$.  For $\pi\pi$ (or $\pi$p)
pairs with $Q >$ 20 MeV/c, WKB is reasonable for $r$ down to $\sim$ 5 fm (or
$\sim$ 6 fm).} Outside the classical turning point the s-wave is
\beq
\psi_C^s(r) \simeq \frac{1}{rk(r)^{1/2}Q^{1/2}}\sin \phi(r),
\label{psiint}
\eeq
where the local relative momentum, measuring the rate of change of phase,
$\phi$, of the wave function, is given by
\beq
k(r) = \frac{d\phi}{dr} =\left(Q^2 \mp \frac{2m_{\rm red}e^2}{r}\right)^{1/2}.
\label{pr}
\eeq
(Equation (\ref{psiint}), with $\ell$-dependent $\phi(r)$ holds as well
for higher partial waves, $\ell > 0 $.)  The normalization of (\ref{psiint})
agrees with (\ref{psi0}) as $r\to 0$, while as $r\to\infty$, the Coulomb wave
function behaves as
\beq
\psi_C^s(r) \to \frac{1}{Qr}\sin(Qr -\eta\ln 2Qr +\delta_0).
\label{psiinf}
\eeq

    The distribution of s-wave pairs in the absence of Coulomb interactions is
\beq
\eps_p\eps_p'\left(\frac{d^6n}{d^3pd^3p'}\right)^s_0
= \frac{1}{4} \int dx dx'dx'' dx''' e^{iP(x+ x''-x' -x''')/2}
\nonumber \\
\hspace{80pt}\times \frac{\sin Q|r-r''|}{Q|r-r''|}  \frac{\sin
Q|r'-r'''|}{Q|r''-r'''|}
\langle J^\dagger (x) J^\dagger(x'')J (x''') J(x')\rangle.
\label{jj0s}
%\nonumber \\
\eeq
Then since in the region of any radius $r$ outside the turning point the
Coulomb wave function behaves locally as a free particle s-wave of momentum
$k(r)$, the s-wave pair distribution function is given by
\beq
 \left(\frac{d^6n}{d^3pd^3p'}\right)^s
 \approx \frac{k(r_0)}{Q}
 \left(\frac{d^6n}{d^3pd^3p'}\right)^s_0,
\eeq
where the factor $k(r_0)/Q$ arises from the denominators in Eq.
(\ref{psiint}) and (\ref{jj0s}).  Consequently, $C(q) \simeq
C_0(k(r_0))k(r_0)/Q$, the result in Eq.  (\ref{toydist}) with $Q_0 = k(r_0)$.
Doing classical physics using Coulomb wave functions is again using a steam
roller to crack a nut.

    With the connection between the toy model and the Coulomb wave function we
can now extend the description of Coulomb corrections to smaller values of the
source radius $r_0$.  In general, the effect of the Coulomb interactions
depends on the detailed structure of the source correlation function; let us
describe the localization of the source correlation function $\langle
J^\dagger (x) J^\dagger(x'') J (x''') J(x') \rangle$ in both $|\vec r-\vec
r\,''\,|$ and $|\vec r\,'-\vec r\,'''\,|$ by writing
\beq
\langle J^\dagger (x) J^\dagger(x'') J (x''') J(x') \rangle
\approx s(x-x'') s(x'-x''')
\nonumber \\
\times(\langle J^\dagger (x) J(x') \rangle_0
\langle J^\dagger(x'') J(x''') \rangle_0
%\nonumber \\
+ \langle J^\dagger (x) J(x''') \rangle_0
\langle J^\dagger(x'') J(x') \rangle_0),
\eeq
where $s(x-x')$ defines the effective initial separation of the pair.  For
small relative momenta, the combination $\psi_C(x)s(x)\equiv f(x)$ varies
slowly on a scale of the coherence length $\xi_c$.  Then we find, roughly,
\beq
\eps_p\eps_p'\frac{d^6n}{d^3pd^3p'}&
\nonumber \\
=& \frac{1}{4} \int dx dx'dx'' dx''' f(w)(f^*(w)+f^*(-w))
e^{iP(x+ x''-x' -x''')/2}
%\nonumber \\ &\times
\langle J^\dagger (x) J(x')\rangle_0 \langle J^\dagger(x'') J(x''') \rangle_0,
\label{coulcorr}
\eeq
where $w= (x-x'+x'''-x'')/2$.  The integrals in this equation are
sufficiently involved that the Coulomb corrections would have to be extracted
numerically.  However, if as an approximation we simply replace the term
$f(w)(f(w)+f(-w))$ by its integral over all space, we arrive basically at
Pratt's formula \cite{pratt86}:
\beq
C(Q)\simeq \int d^3r (|f(r)|^2+f(r)f^*(-r)) C_0(Q),
\label{prattcoul}
\eeq
for the modification of the correlation function by Coulomb interactions.

    The correction to the direct term in Eq.~(\ref{prattcoul}) has the form
\beq
C(Q)_{\rm dir}= \int |\psi_C(r)|^2 |s(r)|^2.
\label{direct}
\eeq
To illustrate the transition from the Gamow correction to the toy model
let us take $|s(r)|^2$ to be a normalized Gaussian of range $r_0$:  $|s(r)|^2=
(2\pi)^{-3/2}r_0^{-3} \exp(-r^2/2r_0^2)$.  We show, in Fig. 16, for the
$\pi^+\pi^-$ system, the results of calculations of the correction term $\int
|\psi_C(r)|^2 |s(r)|^2$ for $r_0$ = 1, 5, 9, and 18 fm (dash-dot curves, the
highest for $r_0$ = 1 fm, and falling with increasing $r_0$).  As $r_0\to 0$,
the projection of the square of the Coulomb wave function onto the source
$|s(r)|^2$ converges to the standard Gamow correction (solid line); for $r_0 <
0.1$ fm (not shown in Fig. 16) the difference between the Gamow correction and
a calculation with Eq.~(\ref{direct}) is less than 0.5\%.  For larger $r_0$
values the correction rather quickly approaches the prediction of the toy
model (shown here for an initial radius of 9 fm as a dotted curve), indicating
that, for pairs originating outside their classical turning point, the toy
model provides an adequate and reasonably accurate description of the Coulomb
effects.
\vspace{.5cm}

\begin{figure}
\begin{center}
\epsfig{file=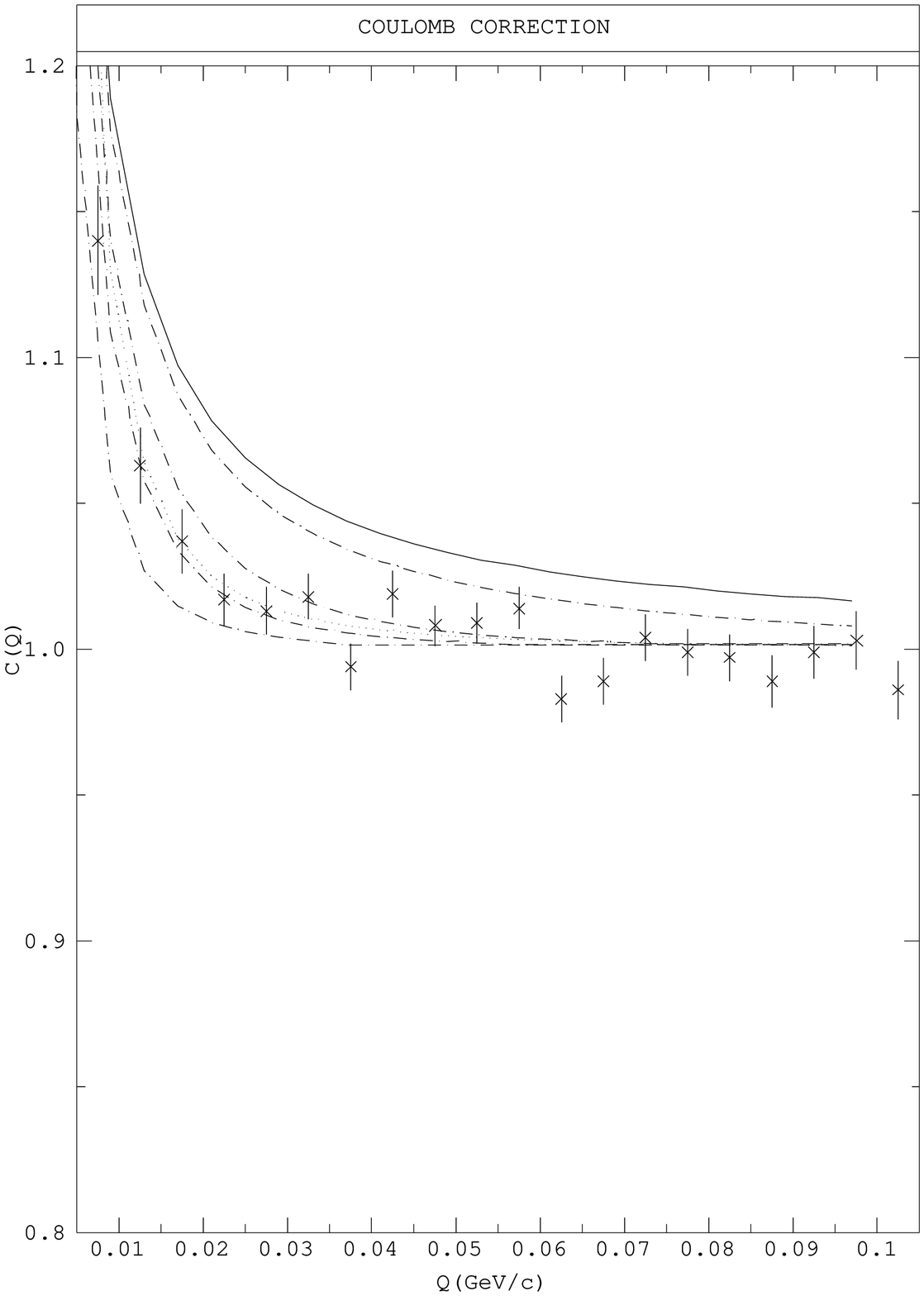, height=14cm}
\end{center}
    Fig.~16.  Transition from the toy model (dotted line, with $r_0$ = 9 fm)
to the Gamow correction (solid line) with decreasing source size, calculated
from Eq.~(\ref{direct}) (dash-dot curves).  From highest to lowest dash-dot
curves the source range $r_0$ is 1, 5, 9, and 18 fm.
\end{figure}

    Let us turn next to the question of the effects of the Coulomb
interactions of the pair with the remaining particles.  This is a difficult
many-body problem, which we greatly simplify as a first approximation by
assuming that the remaining particles can be described by a central Coulomb
potential, $Z_{\rm eff}e^2/r$, where in a central collison of nucleus A with
nucleus B the effective charge $Z_{\rm eff}$ is of order of the total initial
nuclear charge ($Z_A + Z_B$).  This central potential accelerates positive
mesons away and slows down the negatives, effects described by the Coulomb
wave functions for the potential.  The final momentum of any particle is
related to the initial momentum $p_a$ at production point $r_a$ by
\beq
\epsilon(p)= \epsilon(p_a) \pm \frac{Z_{\rm eff}e^2}{r_a},
\label{central}
\eeq
where $\epsilon(p) = (p^2+m^2)^{1/2}$.  (While Coulomb effects for the
relative momentum can be treated non-relativistically as in Eq.  (\ref{toy}),
the individual momenta are generally relativistic.)  For simplicity let us
ignore quantum mechanical suppressions or enhancements of the amplitude for
particle emission, as well as possible effects of angular changes in the
individual particle orbits on the particle distributions.  Then the single
particle distribution is modified by the central potential, analogously to Eq.
(\ref{transf}), by
\beq
\frac{d^3n(\vec p\,)}{dp^3} = \frac{d^3n_0(\vec
p_a\,)}{d^3p_a}\frac{d^3p_a}{d^3p} = \frac{p_a\epsilon(p_a)}{p\epsilon(p)}
\frac{d^3n_0(\vec p_a\,)}{d^3p_a}.
\label{singles2}
\eeq
Both the magnitude of the distribution as well as its argument are shifted.
Experimental observation of these effects is reported in Ref. \cite{NA44coul}.

    Although the central potential shifts the singles distribution, it cannot
introduce any correlations among emitted particles that have no initial
correlation in the absence of the central potential, e.g., as one usually
assumes for different species or oppositely charged pions.  If in the absence
of the central potential, uncorrelated particles [$C(Q)=1$] are emitted in
independent free particle states, then in the presence of the potential they
are emitted in Coulomb states for the central potential, but still $d^6n(\vec
p\,,\vec p\,^\prime)/d^3pd^3p' = (d^3n(\vec p\,)/d^3p)(d^3n(\vec
p\,^\prime)/d^3p')$ and $C(Q)$ remains unity.

    For particles that are initially correlated as a consequence of
Bose-Einstein statistics, $d^6n(\vec p\,,\vec p\,^\prime)/dp^3dp'^3$ and
$(d^3n(\vec p\,)/dp^3)(d^3n(\vec p\,^\prime)/dp'^3)$ will be modified both by
the Jacobians of the transformations from initial to final momenta, and shifts
of argument.  However, in forming $C(q)$, the effects of the Jacobians in the
numerator and denominator essentially cancel, and the primary effect is the
shift in the arguments:
\beq
C(\vec q\,) = \frac{\left\{d^6n_2(\vec p_a\,,\vec
p_a\,^\prime)d^3p_a d^3p_a'\right\}} {\left\{(d^3n(\vec
p_a\,)/d^3p_a)(d^3n(\vec p_a\,^\prime)/d^3p_a')\right\}}.
\label{centcorr}
\eeq
Since positive particles are accelerated, the final momentum difference,
$\vec q = \vec p\, - \vec p\,^\prime$, of a positive pair will generally be
larger in magnitude than it is initially, while for negative pairs the final
momentum difference will generally be smaller.  Thus we expect the central
Coulomb potential to cause the size of the collision volume extracted from
positive pairs to be smaller than the actual size, and that from negative
pairs to be larger than the actual size.  As an illustration consider a pair
of relativistic particles whose initial momenta $\vec p_a$ and $\vec
p_a\,^\prime$ are equal in magnitude to $p_a$, and final momenta $\vec p$ and
$\vec p\,^\prime$ equal in magnitude to $p$; then
\beq
q = (p/p_a)q_a \simeq
q_a \left(1 \pm \frac{Z_{\rm eff}e^2/r_a}{p_a}\right),
\label{expand}
\eeq
where the upper sign refers to both particles positively charged and the
lower to both negatively charged.  For $Z\sim 150$, $r_a \sim 7$ fm and $p_a
\sim$ 300 MeV/c, the effect is an increase for positives (and a decrease for
negatives) in the observed scale of $C(Q)$ and decrease (or increase) in the
extracted radius of ten percent.  Such a shift of the same magnitude has
been observed by E877 in 10.8 GeV/A collisions of Au on Au \cite{e877a};
however, NA44 recently reports an effect in the opposite direction, in 158
Gev/A Pb on Pb collisions, in radii as a function of charged particle
multiplicity \cite{sakaguchi}, indicating the need for a more refined theory
of the effect of the central Coulomb potential \cite{hardtke}.

\section{Applications in condensed matter and atomic physics}

    Let me finally mention briefly work on HBT in condensed matter and atomic
physics.  Recently, Yasuda and Shimizu (at Tokyo University) have made the
first measurement of HBT correlations in an atomic system, observing the time
correlations in laser-cooled ultracold (but not yet Bose-Einstein condensed)
beams of bosonic $^{20}$Ne atoms.  The correlations in the beam are those
expected from a thermal source, where the correlation time is the inverse of
the temperature of the beam.  Indeed the HBT correlation function begins to
rise at time separations less than $\sim 0.5\times10^{-6}$ sec to a value a
factor of two larger than at large time, corresponding to a beam temperature
$\sim 10^2\mu$K.  This result is very similar to the original Hanbury
Brown-Twiss tabletop experiment on photon bunching from a Hg vapor lamp.  By
contrast, a measurement of HBT correlations in the MIT atomic laser
\cite{atomlaser} would yield no such atomic bunching, because of the coherence
of the beam, but rather the correlation function would remain flat.  Lack of
an HBT enhancement would indicate coherence of the beam.  In general, loss of
HBT correlations would probe the onset of Bose-Einstein condensation, not only
in atomic systems, but in condensed matter systems such as the observed
Bose-condensed paraexcitons in cuprous oxide~\cite{wolfe,kbw}.

    Another interesting application of HBT has been in light scattering from
atoms trapped in optical lattices \cite{orsay}.  Jurczak et al.  (at Orsay)
have created an optical lattice with an arrangement of four lasers in which
they trap atomic rubidium at a density $\sim 2\times10^9$ cm$^{-3}$,
filling about $10^{-4}$ of the lattice sites.  The lasers also scatter from
the rubidium, and the time correlations in the scattered light (of two
different polarizations) effectively measure the atom-atom correlation
functions in the lattice.  From these measurements they are able to measure
the diffusion of the loosely packed atoms in the optical lattice.  Lastly we
mention that HBT has also been proposed as a probe of the space and time
structure of bubbles in sonoluminesence \cite{sono}.

    In summary, the technique of Hanbury Brown and Twiss, which was first
developed to measure astronomical object of sizes at least $10^{12}$ cm, has,
as we have seen, turned into a valuable tool to measure subatomic phenomena on
the quite opposite scale of $10^{-12}$ cm.  More recent experiments have shown
its utility in atomic and condensed matter physics as well.  While the basic
theory underlying the nuclear applications is established, as described in
these lectures, many effects, e.g., Coulomb interactions, possible
non-chaoticity, non-zero coherence lengths, multiple scattering, etc.,
introduce various levels of uncertainity into the interpretation of the HBT
measurements.  A better understanding of such effects remains a challenge in
an accurate connection of HBT measurements to the microscopic physics of
collisions.

    These lectures are a small birthday tribute to my dear friend Wies{\l}aw
Czy{\.z} who over the years opened many worlds to me -- from Zakopane and
Cracow, to the pleasures of high energy nuclear physics.  I would like to take
this opportunity to express my gratitude to the organizers of the present
Cracow School of Theoretical Physics at Zakopane for enabling me to
participate in the School where these lectures were given.  I would also like
thank the members of my group in Urbana -- Alejandro Ayala James Popp, and
Benoit Vanderheyden -- who are responsible for much of the material reported
here, and Michael Baym for preparing the graphics.  I am also grateful to
Peter Braun-Munzinger, Henning Heiselberg, Barbara Jacak, and Dariusz
Miskowiec for many discussions of this material and for making figures
available, and to Ulrich Heinz for insightful comments on the manuscript.
This work was supported in part by U.S.  National Science Foundation Grant No.
PHY94-21309.

\end{document}